%% file: NGC-4696-12-09.tex
\documentclass[useAMS,usegraphicx,usenatbib]{mn2e} 
\usepackage{amsmath,graphicx}
\usepackage{subfigure}
\usepackage{ulem}
\usepackage{color}
\usepackage{graphicx}
\usepackage{times} 
\usepackage{amssymb}
\usepackage{amsmath}
\usepackage{lscape}
\usepackage{url}
\usepackage{multirow}
\input{defn}
\newif\ifAMStwofonts
\AMStwofontstrue
\begin{document}

\title[The filaments of NGC 4696] {A deep spectroscopic study of the filamentary nebulosity in NGC\,4696, the brightest cluster galaxy in the Centaurus cluster} \author[R.E.A.Canning et al.] {\parbox[]{6.in}
  { R.~E.~A.~Canning$^{1}$\thanks{E-mail:
      bcanning@ast.cam.ac.uk}, A.~C.~Fabian$^{1}$, R.~M.~Johnstone$^{1}$, J.~S.~Sanders$^{1}$, C.~S.~Crawford$^{1}$, G.~J.~Ferland$^2$ and N.~A.~Hatch$^3$\\ } \\
  \footnotesize
  $^{1}$Institute of Astronomy, Madingley Road, Cambridge, CB3 0HA\\
  $^{2}$Department of Physics, University of Kentucky, Lexington KY 40506, USA\\
  $^{3}$University of Nottingham, School of Physics \& Astronomy, Nottingham NG7 2RD}

\maketitle

\begin{abstract} 
We present results of deep integral field spectroscopy observations using high resolution optical (4150-7200~\AA) VIMOS VLT spectra, of NGC 4696, the dominant galaxy in the Centaurus cluster (Abell 3526). After the Virgo cluster, this is the second nearest (z=0.0104) example of a cool core cluster. NGC 4696 is surrounded by a vast, luminous H$\alpha$ emission line nebula (L$_{\mathrm{H}\alpha}=2.2\times10^{40}$\ergps). We explore the origin and excitation of the emission-line filaments and find their origin consistent with being drawn out, under rising radio bubbles, into the intracluster medium as in other similar systems. Contrary to previous observations we do not observe evidence for shock excitation of the outer filaments. Our optical spectra are consistent with the recent particle heating excitation mechanism of Ferland et al.
\end{abstract}


\section{Introduction}

In this paper we report on deep ($\sim$12 hours) integral field spectroscopy observations of the extended optical emission line system surrounding NGC 4696, the brightest cluster galaxy (BCG) in the Centaurus cluster (Abell 3526), which, at a redshift of $z=0.0104$ is the second nearest example of a `cool core' cluster. 

Nearly half of all galaxy clusters have dense cores of cool X-ray emitting gas observed as a sharp peak in their X-ray surface brightness profiles. These `cool core' clusters have short cooling times and central temperatures which drop to about a third of the cluster ambient temperature (for a review see \citealt{peterson2006} and \citealt{mcnamara2007}). In addition to the sharply peaked X-ray surface brightness profiles, many cool core clusters have extended filamentary optical emission line systems around their central, BCGs (see for example \citealt{hu1985, heckman1989, crawford1992, donahue1992}) and also vast molecular gas and dust reservoirs (e.g. \citealt{edge2001, edge2002, jaffe2001, salome2003, donahue2000, mcnamara1996}). 

Without some form of heating, the hot gas in these short central cooling time clusters will have had time to cool and condense producing large quantities (10$^{11}$-10$^{13}$~\Msun) of cold gas and 1000's of solar masses per year of star formation. Although signatures of cool and cold gas are observed in the central galaxies, in general only $\sim$10 per cent of the predicted star formation is observed. The current paradigm is that some form of heating, most likely dominated by feedback from the central AGN is responsible for regulating the heating and cooling in these objects.  

The optical emission line nebulae surrounding the central galaxies in these systems, have been found to extend to distances of $\sim$100~\kpc\ in some objects and contain $\sim$10$^{4}$-10$^{7}$~\Msun\ of 10$^{4}$~K gas. The origin and excitation mechanisms of the extended optical emission line systems has been much studied but as of yet no one mechanism has been found that can explain in detail all the features of the emission.

Mechanisms proposed include photoionisation by the central AGN (e.g. \citealt{heckman1989}), photoionisation by massive or extremely hot stars (e.g. \citealt{terlevich1985, johnstone1987}), heating through shocks (e.g. \citealt{cowie1980, sabra2000}), conduction or mixing layers in the hot intracluster medium (ICM, e.g. \citealt{sparks1989, crawford1992}), and heating of the filaments through suprathermal particles \citep{ferland2008, ferland2009}.

\begin{table*}
\centering
 \begin{tabular}[h]{|c|c|c|c|c|c|c|}
   \hline
   \hline
   Grism & Order sorting filter & Wavelength Coverage & Exposure & FOV & Spectral Resolution & Dispersion \\
   \hline
         &                      & $\AA$               & s        &     &                     & $\mathrm{\AA}$/pix  \\
   \hline 
   HRO$^{1}$   & GG435                & 5250-7400           & 35100 & 27''$\times$27'' & 2650                & 0.6        \\
   HRO$^{2}$   & GG435                & 5250-7400           & 5400 & 27''$\times$27'' & 2650                & 0.6        \\
   HRB   & free      & 4150-6200           &   9000       & 27''$\times$27'' & 2550                & 0.51       \\
   LRB   & OS blue   & 4000-6700           &   2700       & 54''$\times$54'' & 220                 & 5.3        \\
   \hline
 \end{tabular}
\caption{Observation log. \newline
 $^{1}$ Observations centered on NGC 4696. \newline
 $^{2}$ Scan through galaxy. \label{data_table} }
\end{table*}

The heating and cooling in the Centaurus cluster is very well balanced despite the short central cooling time of only 200~\Myr. NGC 4696 houses a radio source, and multiple bubbles of relativistic gas. These are accompanied by soft X-ray filaments and a sharp rise in the metal abundance in the central 30~\kpc, among the highest seen in any cluster ($\sim$~twice solar, \citealt{sanders2002, fabian2005b, sanders2006}). Centaurus also has one of the broadest range of measured X-ray temperatures, containing gas from 0.35 to 3.7~\keV, over a factor of 10 in temperature \citep{sanders2008b}.

\begin{figure}
\centering
\includegraphics[width=0.45\textwidth]{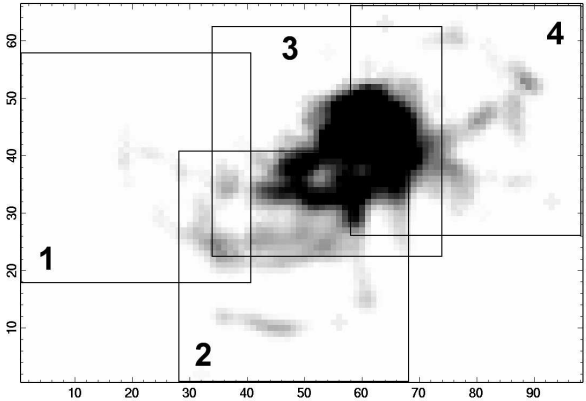}
\caption{Our VIMOS [N {\small II}]$\lambda$6583 emission image of NGC 4696 overlaid with our VIMOS
high resolution pointings. The central RA and Dec. of box 1 is 
$12^{h}48^{m}51^{s}.4$, $-41^{\circ}$18'42.8'', box 2 $12^{h}48^{m}49^{s}.8$, $-41^{\circ}$18'54.3'',
box 3 $12^{h}48^{m}49^{s}.5$, $-41^{\circ}$18'39.8'' and box 4 $12^{h}48^{m}48^{s}.0$, 
$-41^{\circ}$18'37.3'' (all coordinates are in equinox J2000). The x and y axes are in pixels. \label{data}}
\end{figure}

\cite{crawford2005} presented narrow band H$\alpha$+[N {\sc ii}] images showing the extensive, filamentary H$\alpha$ nebulosity surrounding NGC 4696. This shares the morphology of the soft X-ray filaments and of a prominent dust lane. The origin of the complex and extensive filamentary system surrounding NGC 4696 has long been discussed with suggestions that the filaments have cooled from the ICM, are the result of a merger \citep{sparks1989} or have been drawn out into the ICM from the central galaxy by rising radio bubbles \citep{crawford2005}. Recent work by \cite{farage2010} found in support of the merger origin and presented evidence for shock excitation in the filaments. Our data suggest that the filaments are similar to those seen around many other BCGs many of which have not undergone recent merger activity. We uncover a second velocity component in the central region which may be a filament extending behind the galaxy and discuss the excitation mechanisms for the system.

The observations and data reduction are briefly described in $\S$2, analysis, results and a discussion of their implications in $\S$3 and in $\S$4 we summarise our results and main conclusions. At the redshift of the Centaurus cluster ($z=0.0104$, 44.3~\Mpc) one \asec\ corresponds to 0.210~\kpc\ (throughout this paper we adopt H$_0=71$~\kmpspMpc, $\Omega_{\mathrm{M}}=0.27$ and $\Omega_{\Lambda}=0.73$).

\section{Observations and Data Reduction}
\label{obs}

Observations were made on 2009 March 27th-30th using the VIsible MultiObject Spectrograph 
(VIMOS) on the VLT in La Paranal, Chile (see \citealt{lefevre2003} and \citealt{zanichelli2005} for a 
description of the VIMOS IFU and a discussion of data reduction techniques). We obtained high 
resolution orange (HRO), high resolution blue (HRB) and low resolution blue (LRB) data 
using the VIMOS Integral Field Unit (IFU). We used the larger 0.67'' fibres 
giving a field of view of 27''$\times$27'' with the HR grism and 54''$\times$54'' with the LR
grism. Details of the observations are given in Table \ref{data_table} and the HRO pointings
are shown in Fig. \ref{data}.

The deep HRO data cover the wavelength of redshifted spectral lines of [N {\small I}]$\lambda$5199 
to [O {\small II}]$\lambda$7330
including five coronal lines the details of which can be found in \cite{canning2011a}. 
The HRB data cover the wavelength of redshifted lines of H$\gamma~\lambda$4341 to He {\small I} 
$\lambda$5875. An analysis of the stellar spectra in this object using the LRB data will be presented in
Canning et al. 2011 (in prep.).

The data were reduced by the VIPGI\footnote{VIPGI-VIMOS Interactive Pipeline Graphical Interface, 
obtained from http://cosmos.iasf-milano.inaf.it/pandora/.} pipeline \citep{scodeggio2005}. The resulting 3D-datacubes were combined with a set of IDL routines (R. 
Sharp, Private communication).

Instrument flexures in VIMOS are dependent on rotator position and suffer from hysteresis 
\citep{Amico2008}. For this reason we took calibration frames (three flats and
one arc) after each observation block. The observation blocks consisted of three science exposures.
Each were short exposures of only 15 minutes so as to minimise extreme flexures.

\begin{figure}
\centering
\includegraphics[width=0.45\textwidth]{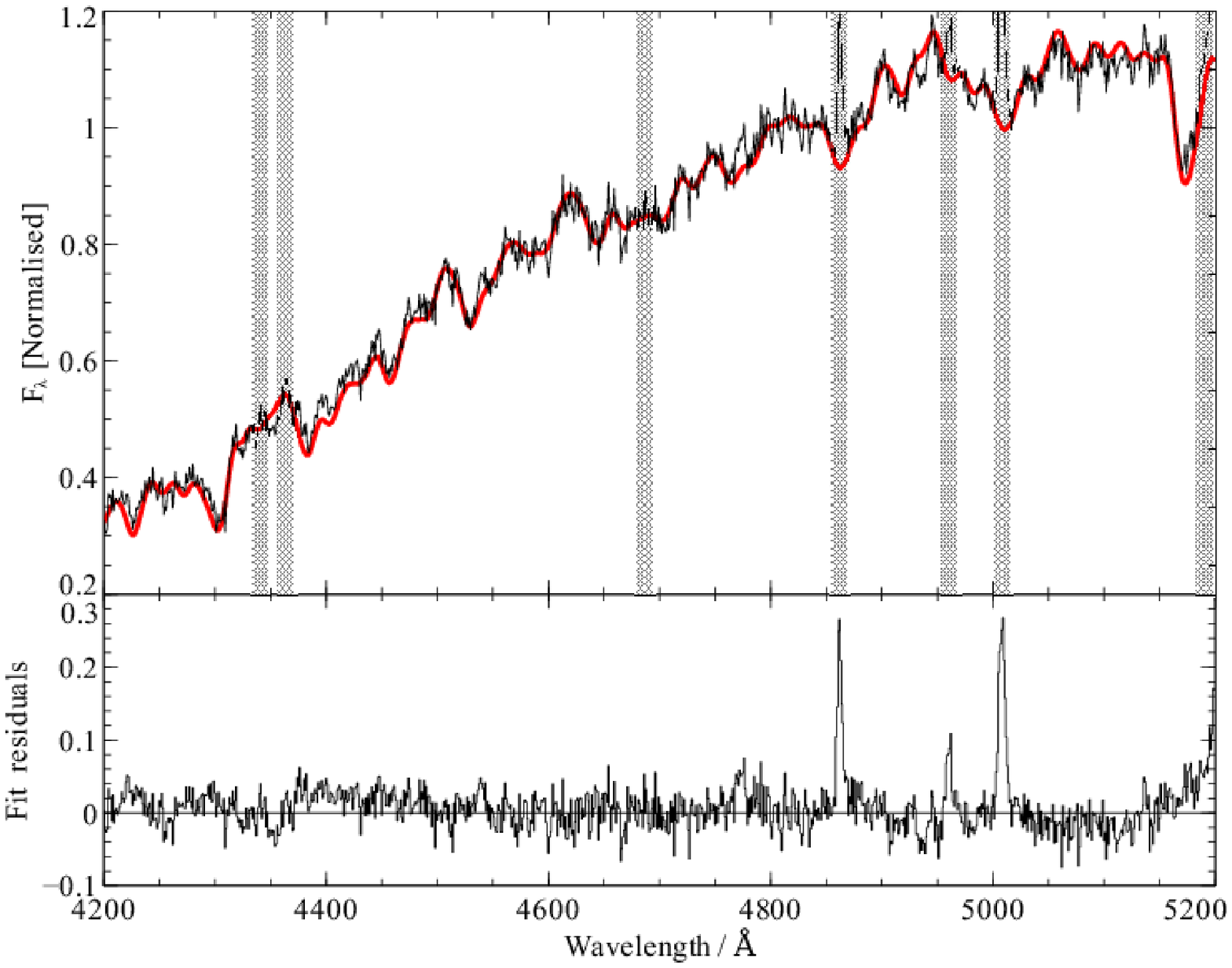}
\caption{An example BC03 SSP model fit (red) to the observed spectrum (black) from 
bin 5 (60,46), between the wavelength of $4000-5200~\mathrm{\AA}$. The fit
residuals are shown in the bottom panel. Any regions where emission lines are expected
are masked out in the fit. In the above plot the masked regions correspond to the wavelengths
of, from left to right, H$\gamma$ $\lambda$4341, [O {\small III}]$\lambda$4363,
He {\small II} $\lambda$4686, H$\beta$ $\lambda$4861, [O {\small III}]$\lambda$4958,
[O {\small III}]$\lambda$5007 and [N {\small I}]$\lambda$5199. \label{SSP}}
\end{figure}

\begin{figure*}
\centering
\includegraphics[width=0.45\textwidth]{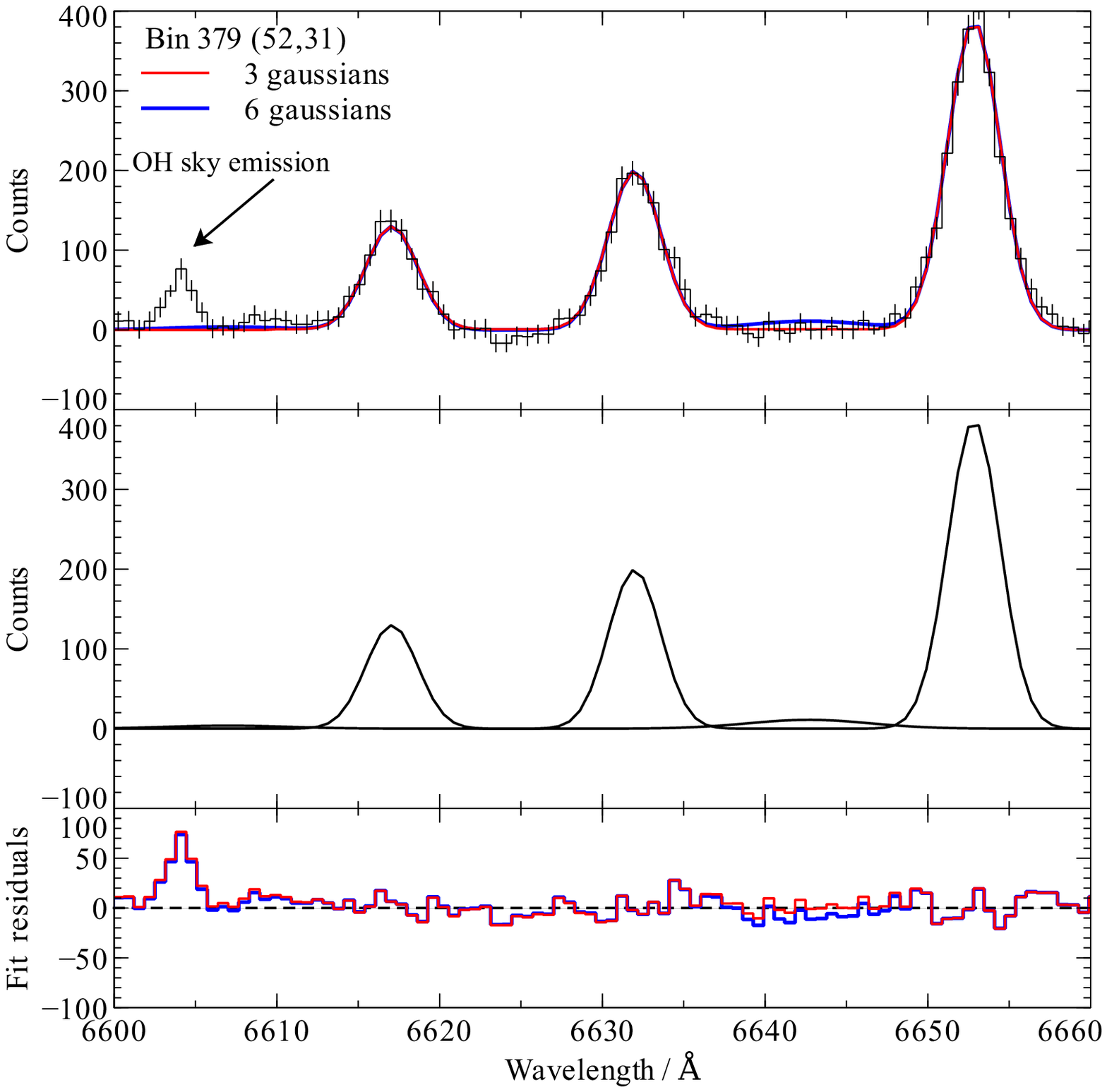}
\includegraphics[width=0.45\textwidth]{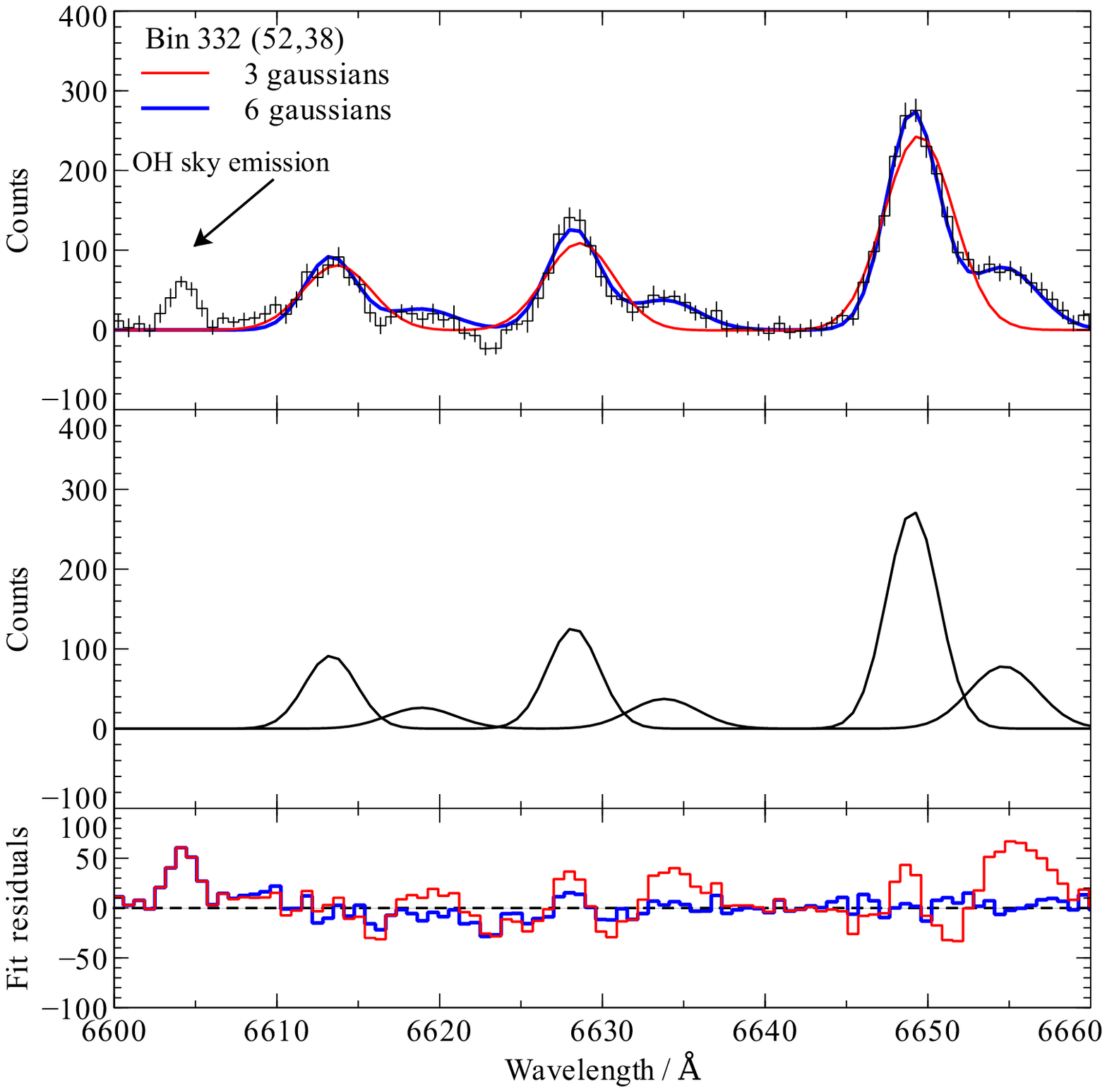}
\caption{Example spectra with one and two velocity component fits from two pixels south
east of the centre of NGC 4696, the pixel coordinates in (x,y) and corresponding 'bin' id is shown in the top left hand corner of the plot. Both these spectra are within the spatial region where we have our 
deepest data. The top panel shows the data (black) and model (single gaussian in red, double gaussian in blue), the middle panel shows the separate gaussian components of the two velocity component model and the lower panel the fit residuals as data$-$model; red residuals are data$-$one velocity component, blue are data$-$two velocity components model. \label{ftest1}}
\end{figure*}

The wavelength calibration was done within VIPGI using He, Ne and Ar arc frames taken throughout the night. VIPGI performs a polynomial fit to the arc lines and we found that the rms deviations are approximately gaussian with a mean of $\sim$0.05$~\mathrm{\AA}$ and dispersion of $\sim$0.02$~\mathrm{\AA}$. This translates to an error in the velocity of $\sim$5~\kmps\ which is significantly smaller than our FWHM instrumental resolution of 98\kmps ($\sigma=42$\kmps). 

Flux calibration was done using standard stars observed at the beginning and end of every night. 
Cosmic-ray rejection, final fibre-to-fibre transmission corrections, sky subtraction,
extinction corrections and shifting to the object rest frame
were performed outside of VIPGI using IDL routines. Sky-subtraction, where done, was 
complicated by the lack of suitable sky fibres and is described below.

Telluric absorption feature corrections for the $\mathrm{O_{2}}$ and $\mathrm{H_{2}O}$ absorption 
in the $6000-7000\,\mathrm{\AA}$ regime was determined from 4 observations of standard stars. Standards were observed at 
the beginning and the end of the night. Details of the correction are given in Appendix \ref{appendixa}.

\begin{figure*}
 \centering
\includegraphics[width=0.85\textwidth]{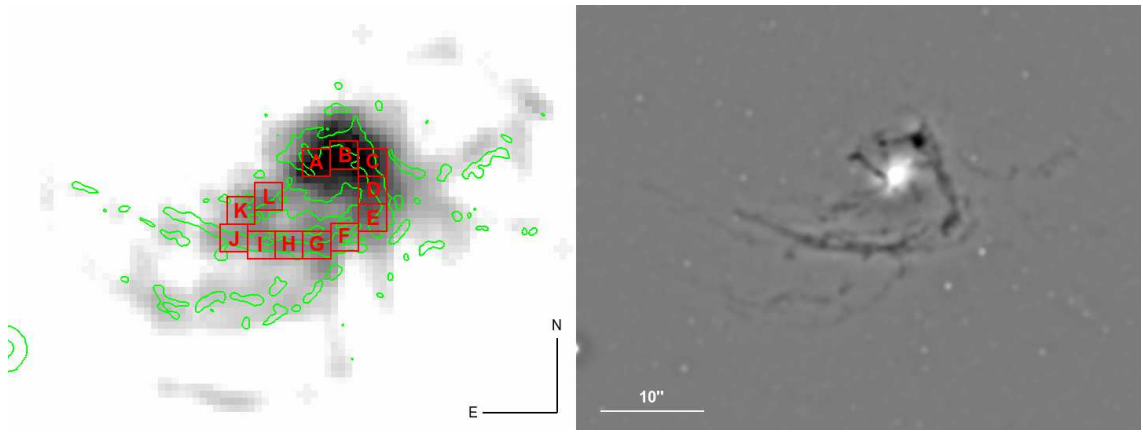}
\caption{The left hand image shows [N {\small II}]$\lambda$6583 emission. The bright ridge of emission which curls round the south-west of the nucleus is clearly visible. The ridge is binned in 2.5 arcsecond squared regions and labelled A to L with A starting near the nucleus. Overlaid are contours from a unsharpmasked broadband HST F435w image (shown on the right, both images have the same scale) tracing the dust lanes. The HST image has been smoothed by 20 pixels and 5 pixels and the two resulting images subtracted. \label{inner_and_dust}}
\end{figure*}

VIPGI's sky subtraction technique is optimised for deep survey observations 
where the field is devoid of extended objects.
Given this we elect not to use VIPGI's sky subtraction technique and instead use 
specific fibres shown to be lacking in emission lines associated with NGC 4696. 
We median combine the spectra to find an average sky spectrum, remove the 
continuum by means of a spline fit and scale and subtract the sky spectra from
our object fibres. The line profiles are slightly different in each quadrant so 
this process is performed on a quadrant by quadrant basis.

After basic reduction, transmission correction and corrections for sky 
absorption and emission we bin the spectra in order to reach a chosen signal
to noise. We try two binning methods: the first bins on the basis 
of the surface brightness of emission in a 2D image by following contours
of a smoothed image \citep{sanders2006c} while the second uses Voronoi
tessellations to provide compact spatial bins \citep{cappellari2003}. The
results of the two binning methods applied to a 2D image of 
[N {\small II}]$\lambda$6583 emission in NGC 4696 is shown in Fig. \ref{binning}.

When binning spectroscopic data one has to be careful not to bin over regions which
are too extended. The spectral properties are unlikely to be similar in
two distinct regions of a galaxy and thus binning these regions together would
lead to spurious results. The Voronoi tessellation method of 
\cite{cappellari2003} provides more compact bins, however the extended optical
filament system of BCGs adds a further complication. These structures are 
extended in only one direction and as such the requirement for `round' bins 
does not trace their morphology particularly well. Due to the filamentary structure
of the optical line nebulosity surrounding NGC 4696 we choose to bin our spectra
using the contour binning technique of \cite{sanders2006c}.

Our HRO and HRB spectra overlap between $5000-6000~\mathrm{\AA}$. In order to incorporate
all our data we split the cubes into three wavelength regions. The regions chosen are
$4000-5100~\mathrm{\AA}$, $5100-5800~\mathrm{\AA}$ and $5800-7500~\mathrm{\AA}$. The 
two HR grisms used have a slightly different spectral resolution (see Table \ref{data_table})
with the HRO being marginally larger than that of the HRB grism. In order to combine
both data sets it is necessary to re-bin the HRO spectra, details of the rebinning technique can be found in Appendix \ref{appendixa}.

\begin{figure*}
 \centering
 \includegraphics[width=0.4\textwidth]{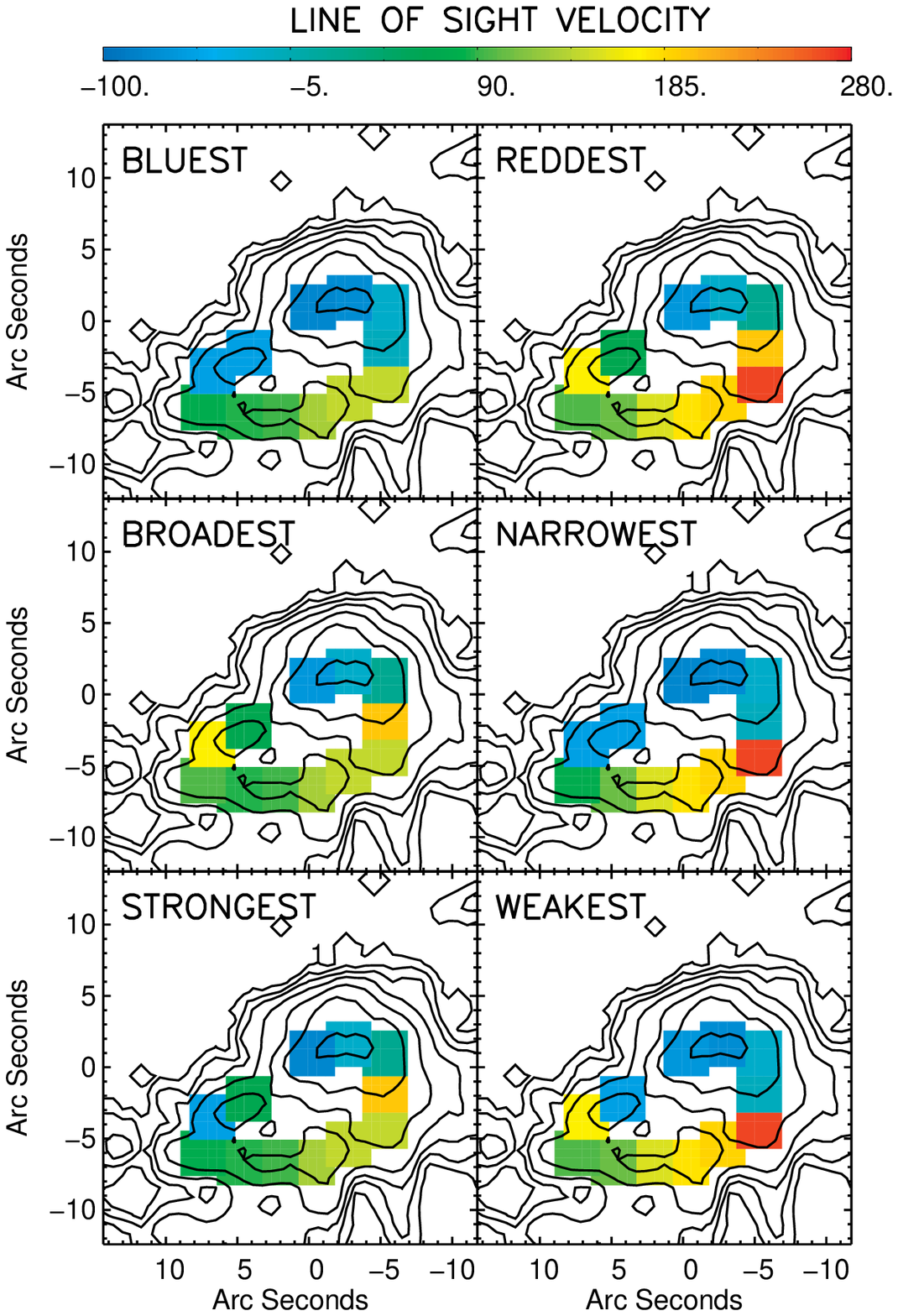}
 \includegraphics[width=0.4\textwidth]{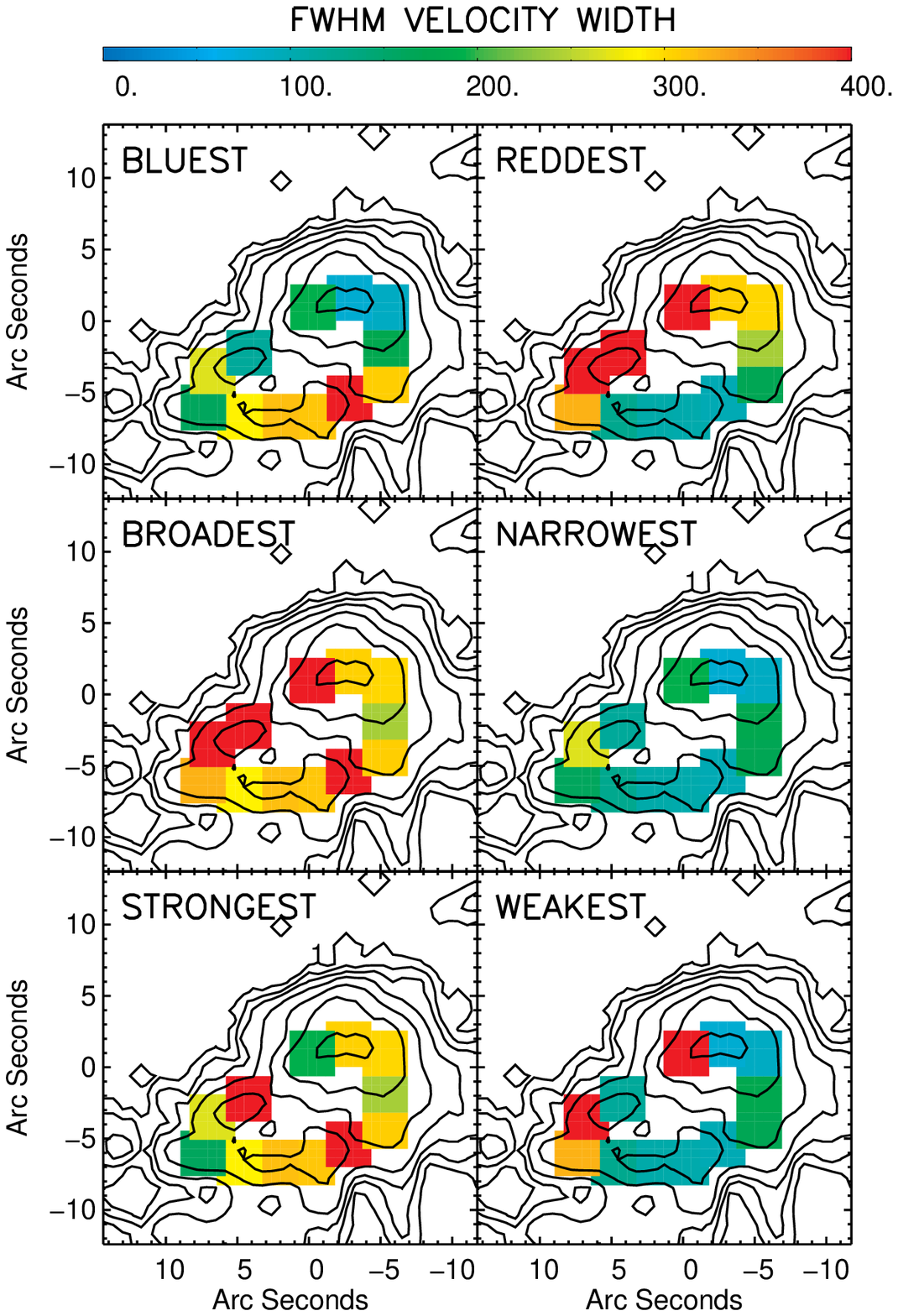}
 \caption{To establish a physical basis for distinguishing the two velocity components detected we investigate the kinematic properties were the two components of the gas to be separated by redshift, velocity dispersion or intensity of emission along the bright inner filament. The figures show the most blueshifted (top left panel), most redshifted (top right), broadest (middle left), narrowest (middle right), strongest (bottom left) and weakest (bottom right) components. Left: The line of sight velocity in \kmps\ of each component with respect to the velocity of NGC 4696 (3045~\kmps). Right: The FWHM velocity width in \kmps\ of each velocity component. The (0,0) position corresponds to the nucleus at a RA and Dec. $12^{h}48^{m}49^{s}.28$, $-41^{\circ}$18'39.4''.
 \label{two_comp}}
\end{figure*}

The strong emission lines of H$\alpha$ and the [N {\small II}] and [S {\small II}]
doublets are fit directly, on a pixel-to-pixel basis across the field of view. We 
fit these in IDL using MPFIT \citep{more1978, markwardt2009}. The weaker
lines are first continuum subtracted using template simple stellar population
(SSP) models from \cite{bc03} (hereafter BC03), fit using the \Starlight\ software package
\citep{cidfernandes2005, cidfernandes2009}. We use basefiles made up of 240 
BC03 models spanning 6 metallicities, covering the range $0.005-2.5~Z_{\odot}$, 
and 40 ages, covering the range $0-20$~Gyr. An example of a fit to the stellar
continuum between 4000 $\mathrm{\AA}$ and 5200 $\mathrm{\AA}$ is shown in Fig. 
\ref{SSP}. After continuum subtraction the emission lines are fit using the 
same technique as the stronger lines described below. Where appropriate we 
impose the kinematics determined from the strongest emission features.

The five strongest emission features ([N {\small II}]$\lambda$6548, H$\alpha\lambda$6563,
[N {\small II}]$\lambda$6583, [S {\small II}]$\lambda$6717 and [S {\small II}]$\lambda$6730) 
are fit simultaneously. The redshift and velocity dispersion are constrained 
to be the same for one velocity component and the integrated flux of the
[N {\small II}] doublet is tied, the scaling being dictated by the atomic
parameters \citep{osterbrock2006}. The spectrum is fit between 6590~$\mathrm{\AA}$ 
and 6820~$\mathrm{\AA}$. A continuum estimate is taken as the region between 
6700$-$6750~$\mathrm{\AA}$ and subtracted from the spectrum. The continuum is fit 
locally for the H$\alpha+$[N {\small II}] emission and the [S {\small II}] doublet.
An example of the fitting technique is shown in Fig. \ref{fit1}. We fit gaussians
to the spectra with 1, 2, and 3 velocity components both with and without a broad 
H$\alpha$ component. We perform an F-test to determine which fit to use, this is 
discussed further in section \ref{analysis}.

\section{Results and discussion}
\label{analysis}

The line emission in the central regions of NGC 4696 is complex with many
spectra requiring two velocity components to adequately fit the observed spectrum.
We use the F-test statistic to determine which fit to use in the spectral modelling
of a particular fibre. The F-test allows us to investigate the probability that 
the data follows the simpler of two proposed models nested within each other. We use 
MPFTEST in IDL to calculate the significance of the addition of extra parameters
in the fit. We choose to accept the additional parameters when the probability falls 
below 0.01. The fits were examined by eye to ensure they were reasonable. Fig. 
\ref{ftest1} shows example fits to two bins; in the first the F-test indicates a single
gaussian fit is sufficient to explain the data while in the second an additional 
velocity component is required.

It was
not found necessary to fit a broad H$\alpha$ component. There is good agreement
between the components found necessary using different binning techniques in all areas
where the data is of a sufficient signal to noise to attain a good fit. We adopt, for 
the remainder of the paper, the contour binning method of \cite{sanders2006c} and 
the appropriate F-test results (see Fig. \ref{ftest2}).

The central region of NGC 4696 is exceedingly complex with many individual filaments, each of which are likely formed by narrower interwoven threads as in the case of the extended emission line system of NGC 1275 in the Perseus cluster \citep{fabian2008}. The two velocity components detected in this central region are probably due to distinct filaments seen in projection and as such it is not obvious that splitting the emission into simple `blueshifted' and `redshifted' populations will lend any physical insight into this system.

\begin{figure}
\centering
\includegraphics[width=0.45\textwidth]{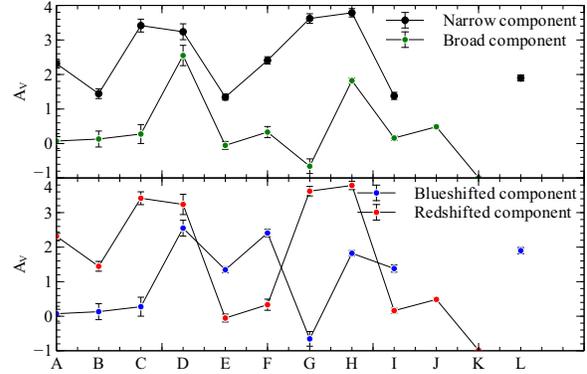}
\caption{The intrinsic A$_{V}$ extinction in the two velocity components derived from their H$\alpha$/H$\beta$ ratio. The top panel shows the extinction were the components to be distinguished as a `broader' and `narrower' component, the bottom panel as a `bluer' and `redder' component. This figure shows splitting the two components based on the width of the line as opposed to the line centroid is equivalent to splitting the components based on the extinction of the emission. 
\label{inner_filament}}
\end{figure}

In order to provide a more physical basis for the division of the two populations we investigate the emission properties along the brightest inner filament. Fig. \ref{inner_and_dust} shows on the same scale the [N {\small II}]$\lambda$6583 emission and dust lanes. There is a bright ridge of emission curling round the south-west of the nucleus which, in projection, coincides exactly with a large dust lane. We bin this region in 2.5$\times$2.5 arcsecond bins (labelled from A to L in Fig. \ref{inner_and_dust}) and investigate the kinematic properties of the two velocity components were they to be seperated based on redshift, velocity width and intensity; the results are presented in Fig. \ref{two_comp}. 

We would expect, if these two components followed emission from two distinct filaments, that the line of sight velocities and the FWHM velocity widths would vary smoothly with distance along the filament. By which we mean that the velocities are unlikely to jump dramatically from bin to bin along the filament length. From this argument and using the kinematical data we rule out a division based on intensity (the bottom panel of Fig. \ref{two_comp}). To investigate further separating the components based on redshift and velocity width we derive the intrinsic extinction from the H$\alpha$/H$\beta$ ratio of each component. The results can be seen in Fig. \ref{inner_filament} and imply that splitting the two velocity components based on the breadth of the emission is equivalent to distinguishing them based on intrinsic extinction, with the narrower component being more extinguished than the broader component in all regions where the H$\alpha$ and H$\beta$ lines could be detected at greater than 4 sigma. Splitting the components based on redshift provides no such distinction.
From these results we suggest that distinguishing the two components by velocity width as opposed to redshift is the more physical approach. The broad component lies closer to the nucleus, coincident with the dust lane, whereas the narrow component lies behind, and is therefore more reddened. For the remainder of the paper, unless otherwise stated, the two velocity components will be split into `broader' and `narrower' width components.

\subsection{Reddening}

We deredden the spectra for Galactic extinction using the reddening law of 
\cite{calzetti2000} and an E(B-V) value for the stellar continuum of 0.11 
($A_{V}=0.34$). An 
extinction to reddening ratio, $R=A_{V}/E(B-V)$, of 3.1 is assumed for our Galaxy.
This is done before shifting the spectra to their rest wavelength.
A value for the intrinsic reddening is also estimated, where possible, from 
the H$\alpha$ and H$\beta$ line ratios (Fig. \ref{reddening}).

The logarithmic H$\beta$ extinction coefficient, $c$, is calculated
from the observed $I_{\lambda}/I_{H\beta}$, and intrinsic line ratios 
$I_{\lambda0}/I_{H\beta0}$, 

\begin{equation}
 \frac{I_{\lambda}}{I_{H\beta}}=\frac{I_{\lambda0}}{I_{H\beta0}}10^{c[f(\lambda)-f(H\beta)]},
 \label{ratio}
\end{equation}

\noindent assuming a reddening law $f(\lambda)$. The extinction coefficient and reddening 
can therefore be calculated through,

\begin{equation}
 A_{\lambda}=-2.5cf(\lambda).
 \label{ext}
\end{equation}

\noindent For the weaker Balmer line we impose the kinematics of the strong emission lines 
and subtract the continuum using template SSP models as described in section 
\ref{obs}. The intrinsic extinction is calculated using the 
rest wavelengths of the lines and we assume a case B intrinsic recombination ratio. This method makes the assumptions that the dust is an obscuring screen, is homogeneous and that all the H$\alpha$ and H$\beta$ emission is from hydrogen recombination, none from collisional excitation.

The reddening map for a one component velocity fit, on a per pixel basis, in the central regions of NGC 4696 is shown in Fig. \ref{reddening}. A single component fit is presented as H$\beta$ is weak and the second velocity component does not provide a good fit unless the fibres are rather coarsely binned, as in Fig. \ref{inner_filament}. Contours of E(B-V) derived from HST B and I band images and \chandra\ X-ray N$_{\mathrm{H}}$ column density \citep{crawford2005} are overlaid. 

The regions where the value of intrinsic A$_{V}$ is highest correspond well to the regions where E(B-V) and N$_{\mathrm{H}}$ are greatest though the exact position of the peaks are sometimes offset from those in the X-ray and broad band optical map. \cite{crawford2005} find the regions of greatest extinction have a value of A$_{V}\sim$0.4 which is in good agreement with previous studies \citep{jorgensen1983, sparks1989}. Our results, for the same knots, assuming a case B intrinsic H$\alpha$ to H$\beta$ line ratio \citep{osterbrock2006} are larger than this with A$_{V}$'s in the range 1.5-2 for most of the knots along the dust lane. In order to make our results consistent with these previous imaging studies we require an intrinsic H$\alpha$ to H$\beta$ ratio of $\sim$4 (nearly 1.5$\times$ the case B ratio) requiring a large component of collisional excitation to the hydrogen lines. 

For Seyfert galaxies, radio galaxies and LINERs the intrinsic H$\alpha$/H$\beta$ ratio is usually measured to be larger than case B, $\sim$3.1, due to the greater importance of collisional excitation in regions where there is a harder photoionising spectrum (e.g. \citealt{veilleux1987}). High ratios have also been seen in the case of the extended emission line regions surrounding some BCGs. \cite{kent1979} found a ratio of 4.77 in the filaments of NGC 1275, the brightest cluster galaxy in the Perseus cluster. Gemini spectroscopy by \cite{hatch2006} confirms this high ratio finding values of H$\alpha$/H$\beta$ of 3-7 in the filaments. \cite{voit1997} also find significant reddening in the filaments of Abell 2597 at 1.5 times the case B ratio. In both NGC 1275 and Abell 2597 the authors find no evidence for a deviation from the Galactic extinction laws. This implies the properties of the dust in the filaments are similar to that of our own Galaxy.

\begin{figure}
\centering
\includegraphics[width=0.45\textwidth]{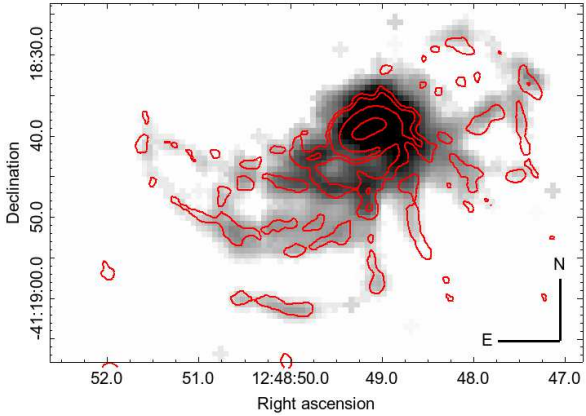}
\caption{[N {\small II}]$\lambda$6583 emission (grey scale) overlaid with contours of H$\alpha$ from the narrow band imaging of \protect \cite{crawford2005}. The morphologies of other emission lines follow generally that of the [N {\small II}]$\lambda$6583 emission. \label{ha}}
\end{figure}

\begin{figure}
\centering
\includegraphics[width=0.44\textwidth]{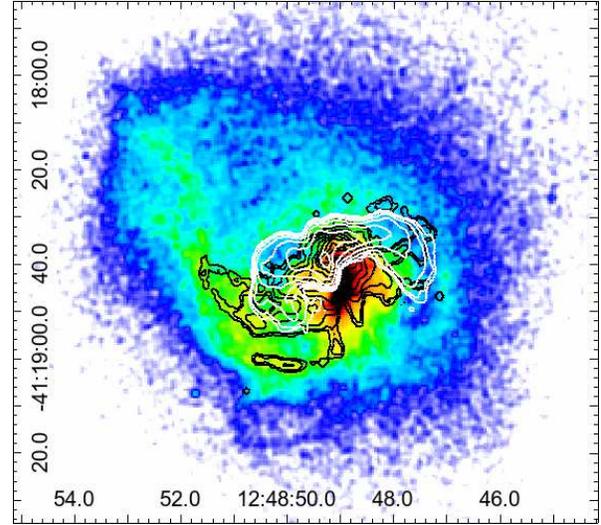}
\caption{Soft X-ray emission (0.5-1.5~keV) of the centre of NGC 4696. Contours of [N {\small II}]$\lambda$6583 emission are overlaid in black and 5~GHz radio contours in white. \label{multi_wav}}
\end{figure}

\cite{sparks1989} found, from broad band V and R images, that the dust extinction with wavelength behaves `normally' in NGC 4696, that is there is no evidence for a departure from the Galactic extinction laws. This makes formation of the filaments from the hot gas unlikely, and the authors favour a merger origin for the dust lane and filaments. More recently it has been suggested that the dust lane could be the result of previous dusty star formation in the central galaxy which has been drawn out with the cool gas into filaments by radio bubbles rising in the ICM.

\begin{figure*}
\centering
\includegraphics[width=0.8\textwidth]{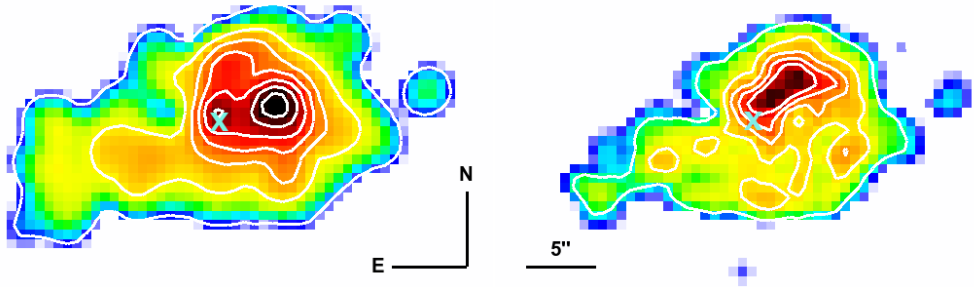}
\caption{The morphologies of the broader (left) and narrower (right) velocity components. The peak in the emission between the two components is offset by $\sim$2''. The broader component has a much more circular profile while the narrow component has a steeper central peak and exhibits a less smooth structure. The two images have been smoothed by a gaussian with FWHM 3.1'' ($\sigma=$1.3'') and are plotted on the same spatial scale; the cyan 'X' marks the nucleus. The intensity scale has been adjusted individually to achieve the best contrast. \label{morph}}
\end{figure*}

\begin{figure*}
\centering
\includegraphics[width=0.8\textwidth]{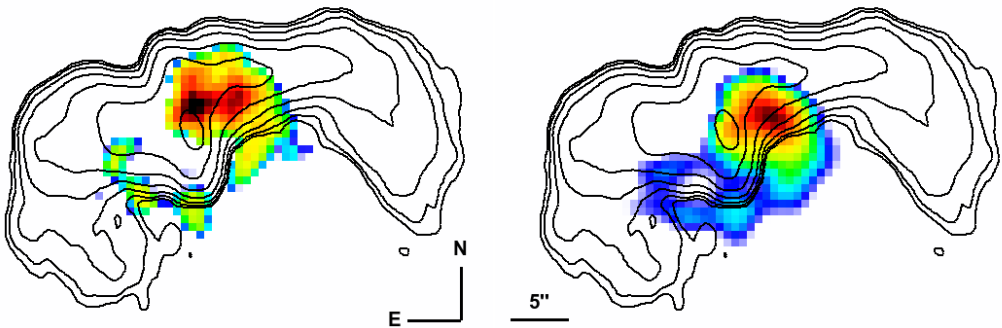}
\caption{The [O {\small III}]$\lambda$5007 (left) and [N {\small II}]$\lambda$6583 (right) emission in the central region of the galaxy. The peak in emission is offset by $\sim$4'' between the two images.  The two images have been smoothed by a gaussian with FWHM 3.1'' ($\sigma=$1.3'') and are plotted on the same spatial scale. The intensity scale has been adjusted individually to achieve the best contrast. \label{morph_oiii}}
\end{figure*}

\begin{figure*}
\centering
\includegraphics[width=0.99\textwidth]{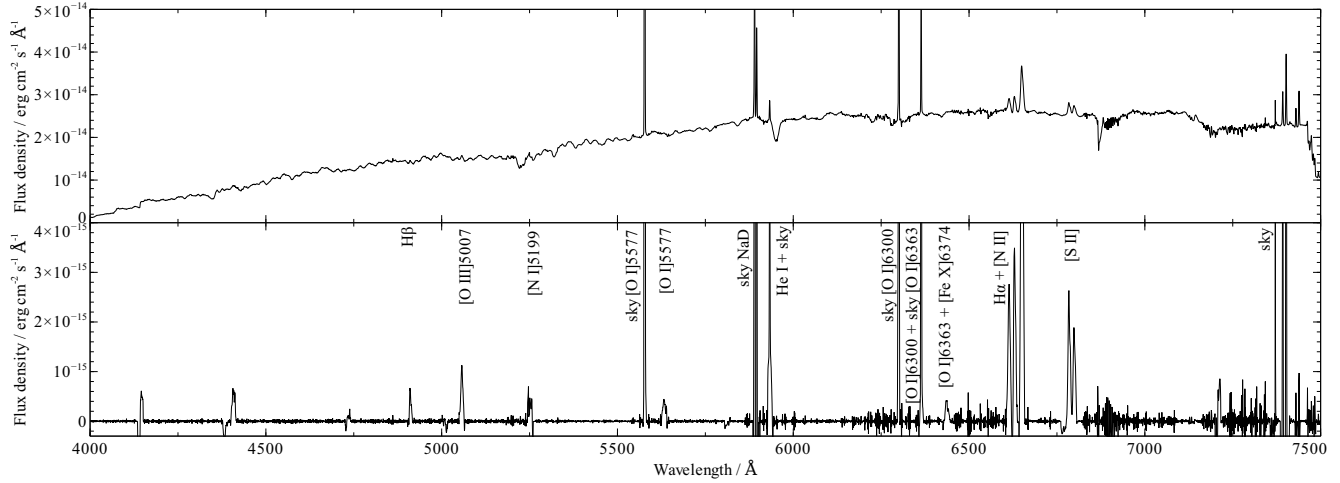}
\caption{Top: The spectrum of NGC 4696 within a radius of 10 arcseconds, sky lines that are confused with object emission are not subtracted. Bottom: The same spectrum after continuum subtraction using BC03 models. Object emission lines and confused sky lines are labelled. \label{spectrum}}
\end{figure*}

\begin{table*}
\centering
 \begin{tabular}[h]{|c|c|c|c|}
   \hline
   \hline
    Line & Total flux & Total luminosity & Flux / F$_{\mathrm{H}\alpha}$ \\
     & \ergpcmsqps & \ergps  &  \\
  \hline
H$\gamma~\lambda   4341$ & $    <2.2\times10^{-15}    $ & $    <4.9\times10^{38}    $&       0.02 \\
$\mathrm{[O III]}~\lambda   4363$ & $    <1.5\times10^{-15}    $ & $    <3.3\times10^{38}    $&       0.01 \\
He~$\mathrm{II}~\lambda   4686$ & $    <2.4\times10^{-15}    $ & $    <5.5\times10^{38}    $&       0.02 \\
H$\beta~\lambda   4861$ & $    2.0\times10^{-14}\pm    9.9\times10^{-16}$ & $    4.5\times10^{39}\pm    2.3\times10^{38}$&       0.20 \\
$\mathrm{[O III]}~\lambda   4958$ & $    9.2\times10^{-15}\pm    1.2\times10^{-15}$ & $    2.1\times10^{39}\pm    2.8\times10^{38}$&       0.09 \\
$\mathrm{[O III]}~\lambda   5007$ & $    2.8\times10^{-14}\pm    5.5\times10^{-16}$ & $    6.3\times10^{39}\pm    1.3\times10^{38}$&       0.28 \\
$\mathrm{[N I]}~\lambda   5199$ & $    5.4\times10^{-15}\pm    1.3\times10^{-15}$ & $    1.2\times10^{39}\pm    2.9\times10^{38}$&       0.06 \\
$\mathrm{[C V]}~\lambda   5309$ & $    <6.0\times10^{-16}    $ & $    <1.4\times10^{38}    $&       0.01 \\
$\mathrm{[O I]}~\lambda   5577$ & $    <2.9\times10^{-15}    $ & $    <6.6\times10^{38}    $&       0.03 \\
$\mathrm{[N II]}~\lambda   5755$ & $    <4.2\times10^{-15}    $ & $    <9.6\times10^{38}    $&       0.04 \\
He~$\mathrm{I}~\lambda   5875$** & $    1.6\times10^{-14}\pm    1.4\times10^{-15}$ & $    3.8\times10^{39}\pm    3.3\times10^{38}$&       0.17 \\
$\mathrm{[O I]}~\lambda   6300$** & $    1.7\times10^{-14}\pm    4.0\times10^{-15}$ & $    4.0\times10^{39}\pm    9.0\times10^{38}$&       0.18 \\
$\mathrm{[O I]}~\lambda   6363$** & $    5.7\times10^{-15}\pm    3.0\times10^{-15}$ & $    1.3\times10^{39}\pm    7.0\times10^{38}$&       0.06 \\
$\mathrm{[N II]}~\lambda   6548$ & $    5.4\times10^{-14}\pm    1.3\times10^{-14}$ & $    1.2\times10^{40}\pm    3.0\times10^{39}$&       0.56 \\
H$\alpha~\lambda   6563$** & $    9.8\times10^{-14}\pm    5.9\times10^{-15}$ & $    2.2\times10^{40}\pm    1.4\times10^{39}$&       1.00 \\
$\mathrm{[N II]}~\lambda   6583$ & $    1.8\times10^{-13}\pm    9.5\times10^{-15}$ & $    4.2\times10^{40}\pm    2.2\times10^{39}$&       1.87 \\
$\mathrm{[S II]}~\lambda   6717$ & $    5.5\times10^{-14}\pm    5.7\times10^{-15}$ & $    1.3\times10^{40}\pm    1.3\times10^{39}$&       0.56 \\
$\mathrm{[S II]}~\lambda   6731$ & $    4.5\times10^{-14}\pm    5.2\times10^{-15}$ & $    1.0\times10^{40}\pm    1.2\times10^{39}$&       0.46 \\
$\mathrm{[Ca II]}~\lambda   7291$** & $    <1.7\times10^{-14}    $ & $    <3.9\times10^{39}    $&       0.18 \\
$\mathrm{[O II]}~\lambda   7320$** & $    <3.7\times10^{-14}    $ & $    <8.4\times10^{39}    $&       0.38 \\
$\mathrm{[Ca II]}~\lambda   7325$** & $    <3.7\times10^{-14}    $ & $    <8.5\times10^{39}    $&       0.38 \\
$\mathrm{[O II]}~\lambda   7330$** & $    <2.1\times10^{-14}    $ & $    <4.8\times10^{39}    $&       0.21 \\
   \hline
 \end{tabular}
\caption{Total fluxes and luminosities of emission lines in NGC 4696 after correction for 
extinction using the median $A_{V}=0.9$. Emission lines blueward of 5100$\mathrm{\AA}$ are only measured in the inner 27'' as we do not have HRB data beyond this. Where the lines are marked by $<$ the 
value indicates a 90 per cent upper limit ($\Delta\chi^{2}=2.7$). Where the lines are
marked with ** the object emission was confused with other emission or sky lines and it
was necessary to fit these simultaneously. All values are
determined after subtraction of the continuum. \label{total_emmision} }
\end{table*}

To make a comparison with the extinction corrections of \cite{farage2010} we examine the extinction in a 3 arcsecond diameter region approximately 3'' north-west of the nucleus (taken to be the centre of radio emission and of the X-ray point source $12^{h}48^{m}49^{s}.28$, $-41^{\circ}$18'39.4''), where we find the peak in our H$\alpha$ emission map. The coordinates of our chosen region are ($12^{h}48^{m}49^{s}.1$, $-41^{\circ}$18'37.8''). Fitting the H$\alpha$ and H$\beta$ emission with a single velocity component, as in \cite{farage2010}, we find the intrinsic A$_{V}$ in this region is 0.33. Using the same values of the intrinsic emitted flux ratio ($\mathrm{H}\alpha/\mathrm{H}\beta=3.1$) as these authors gives a total A$_{V}$ of 0.67. Using the case B ratio the total A$_{V}$ is 0.91. Our value differs from that of \cite{farage2010} by 0.07; this is well within the error as a difference in the position of our region of 1.5'' translates to a factor of almost 2 in the H$\alpha$/H$\beta$ ratio.

As the reddening correction varies quite significantly across the field of view, due to the presence of the large dust lane, we elect to use the local value of the H$\alpha$/H$\beta$ ratio where possible to correct our fluxes. Where this is not possible we use the median value of A$_{V}=$0.9.

\subsection{Filament morphology}
\label{fil_morph}

The detected line emission in our data follows closely the emission detected in the narrow band H$\alpha$ imaging of \cite{crawford2005} (Fig. \ref{ha}). A central filament is seen with two velocity components extending 12 arcseconds to the south-east of the nucleus. A further complex system of filaments extending over 25 arcseconds to the north-west, south and east are also detected. Only the north-east appears devoid of filaments. All emission lines, with the exception of [O {\sc iii}], where the lines were strong enough not to impose the kinematics of the H$\alpha$ emission were found to have the same morphology as the H$\alpha$ line suggesting the emission occurs from the same gas. There also appears to be dust lanes present in all filaments except the weak southern-most filament (Fig. \ref{inner_and_dust}).

The emission line gas traces well the inner regions of the soft X-ray filaments (Fig. \ref{multi_wav}). However, the soft X-ray filaments extend much further up to the north-east of the galaxy leading towards a depression in the X-ray gas thought to be a ghost bubble, of plasma from the radio source, risen bouyantly from the nucleus \citep{crawford2005}. The area just north-east of the nucleus, devoid of both optical emission line gas and soft X-ray filaments coincides with a prominant radio bubble.

The two velocity components have a different morphological structure. Fig. \ref{morph} shows the [N {\small II}]$\lambda$6583 line emission in this central region, smoothed by a gaussian with FWHM 3.1'' ($\sigma=$1.3'', 2 pixels). The left hand image is the emission from the broader component and shows two clear peaks in emission. The weaker peak to the east coincides with the centre of radio emission and the weak X-ray point source ($12^{h}48^{m}49^{s}.28$, $-41^{\circ}$18'39.4'', \citealt{taylor2002}); throughout this paper we define the nucleus as this point. 5 arcseconds to the west is the point of largest flux for the broad emission component. The surface brightness contours show the emission about this point is smooth and circular in the central 5 arcseconds before stretching towards the south-east of the nucleus. The right hand image shows the narrow emission component which is far less smooth in morphology. The emission peaks in a bar stretching north-west from the nucleus then curls clockwise round the nucleus. We find the contours of surface brightness of both of these components deviates from that of the dust near to the nucleus as found by \cite{farage2010}.

Fig. \ref{morph_oiii} shows a single component fit to the [O {\small III}]$\lambda$5007 emission (left), beside an [N {\small II}]$\lambda$6583 single component fit emission map (right). The [O {\sc iii}] emission is weakly detected in the inner filament (we do not have a HRB pointing in the outer regions) but very strong near to the centre of radio emission. The second peak of [O {\sc iii}] emission to the west which coicides with the peak in the other emission lines is much weaker, implying much of the [O {\sc iii}] emission in the central regions is associated with the AGN. 

\cite{sparks1989} and \cite{farage2010} suggest NGC 4696 has an intrinsically one sided spiral structure due to the apparent close association of the main dust lane and line emission. While we also observe a very striking close correspondence between the optical filaments and the many dust lanes of NGC 4696 (see Fig. \ref{inner_and_dust}), our detection of at least two velocity components to the optical emission line gas, exhibiting different extinction properties, provides evidence contrary to this suggestion. NGC 4696 is clearly a very complex galaxy surrounded by an intricate filamentary system. The high extinction of the narrow velocity component and its less smooth morphology is interpreted as a filament located behind the main, more centrally situated broader emission line component.

\subsection{Emission line fluxes}

The spectrum of a region with a radius of 10 arcsec centred on the radio core, with and
without continuum subtraction, is shown in Fig. \ref{spectrum}, with the strongest lines
marked. The spectrum is similar to other filament emission in cool core BCGs, showing 
strong low ionisation emission lines. NGC 4696 has stronger [O {\sc iii}]$\lambda$5007 emission 
than found in NGC 1275 or Abell 2597 though the majority of this emission coincides with
the nucleus and is likely associated with the AGN.

We compare the dereddened flux in H$\alpha$ emission to that of \cite{crawford2005}. Using
narrow band H$\alpha$ imaging the authors find a surface brightness in the inner ($\sim$~8''
from the radio core) filaments of $3.5\times10^{-16}$~\ergpcmsqpsparcsecsq\ and in the
fainter outer filaments ($\sim$~27'') of $1.4\times10^{-16}$~\ergpcmsqpsparcsecsq. They 
obtain a lower limit on the total H$\alpha$ luminosity beyond a radius of 3.5'' of 
$1.5\times10^{40}$~\ergps. Our data are consistent with this with the average surface
brightness of the filaments 8'' and 27'' from the radio core being 
$2.2\times10^{-16}$~\ergpcmsqpsparcsecsq\ and $1.3\times10^{-16}$~\ergpcmsqpsparcsecsq\
respectively. We obtain a total extinction corrected H$\alpha$ luminosity beyond a radius of 3.5'' of
$1.7\times10^{40}$~\ergps and a total extinction corrected H$\alpha$ luminosity of $2.2\times10^{40}$~\ergps.
Our derived H$\alpha$ and [N {\sc ii}] fluxes and luminosities are slightly larger than, though
still consistent with, those of \cite{farage2010}. This is expected as our HRO data probes
more of the filament system. 

The total luminosities of all detected lines and the 90 per cent upper limits of emission 
lines not found in our spectra are given in Table \ref{total_emmision}. Emission lines are 
fit as described in Section \ref{obs} although some lines are confused with other 
emission lines or sky and therefore warrant special attention. He {\small I}, 
[O {\small I}]$\lambda$6300, H$\alpha$, [Ca {\small II}] and emission lines of [O {\small II}] 
suffer from confusion with sky emission, in each of these cases the sky emission is
fit simultaneously with the object emission lines. The red wing of the object 
[O {\small I}]$\lambda$6363 emission coincides with the 
[Fe {\small X}]$\lambda$6374 emission line discussed in \cite{canning2011a}.
The flux of this line is linked by atomic physics to the flux of the 
[O {\small I}]$\lambda$6300 emission and we fit these, the sky [O {\small I}] lines
and the [Fe {\small X}]$\lambda$6374 emission line together. We only have HRB data from the central $\sim$27'' of NGC 4696, therefore all emission lines
blueward of 5100$\mathrm\AA$ may be lower limits.

\begin{figure}
\centering
\includegraphics[width=0.45\textwidth]{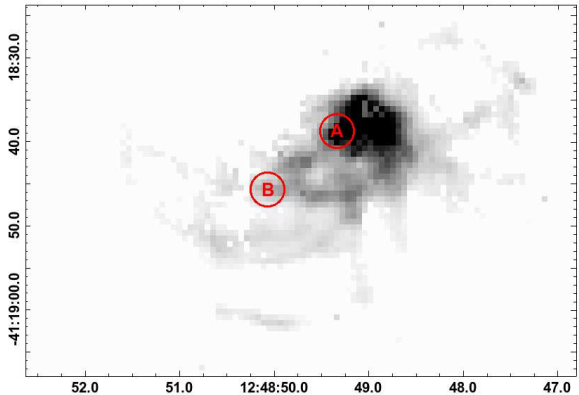}
\caption{Regions A and B, both 3 pixels in radius. Region A is coincident with the nucleus defined as the centre of radio emission ($12^{h}48^{m}49^{s}.28$, $-41^{\circ}$18'39.4'' \protect \citep{taylor2002}, while region B is $\sim$12 arcseconds to the south-east of the nucleus, on the outer edge of the inner filament. \label{inner_outer_circles}}
\end{figure}

\begin{figure}
\centering
\includegraphics[width=0.45\textwidth]{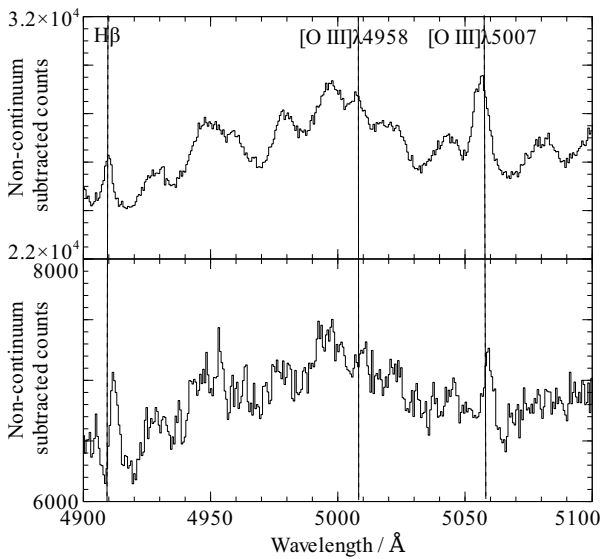}
\caption{The non-continuum subtracted HRB spectra in regions A (top panel) and B (bottom panel). The [O {\small III}]$\lambda$5007/H$\beta~\lambda$4861 ratio varies dramatically from $\sim$2 (region A) to $\sim$0.7 (region B, see table \ref{fluxes}).  \label{oiii_hb_spectra}}
\end{figure}

\begin{table}
\centering
 \begin{tabular}[h]{|c|c|c|}
   \hline
   \hline
    Line &  F$_{\lambda}$/ F$_{\mathrm{H}\alpha~A}$ & F$_{\lambda}$/ F$_{\mathrm{H}\alpha~B}$\\
  \hline
H$\beta~\lambda   4861$           &  0.37       & 0.23\\
$\mathrm{[O III]}~\lambda   4958$ &  0.26       & - \\ 
$\mathrm{[O III]}~\lambda   5007$ &  0.77       & 0.15\\
$\mathrm{[N I]}~\lambda   5199$   &  0.23       & 0.19\\
He~$\mathrm{I}~\lambda   5875$**  &  0.15       & - \\
$\mathrm{[O I]}~\lambda   6300$** &  0.18       & 0.21\\
$\mathrm{[O I]}~\lambda   6363$** &  0.06       & 0.07\\
$\mathrm{[N II]}~\lambda   6548$  &  1.04       & 0.57\\
H$\alpha~\lambda   6563$**        &  1.00       & 1.00\\
$\mathrm{[N II]}~\lambda   6583$  &  3.10       & 1.70\\
$\mathrm{[S II]}~\lambda   6717$  &  0.92       & 0.39\\
$\mathrm{[S II]}~\lambda   6731$  &  0.58       & 0.33\\
   \hline
 \end{tabular}
\caption{Extinction corrected F$_{\lambda}$/ F$_{\mathrm{H}\alpha}$ in regions A and B (see Fig. \ref{inner_outer_circles}). A coincides with the nucleus, defined as the centre of radio emission while B is a point $\sim$12 arcseconds away on the north-eastern edge of the inner filament. Where the lines are
marked with **, the object emission was confused with other emission or sky lines and it
was necessary to fit these simultaneously. The measured H$\alpha$ fluxes in the two regions are: F$_{\mathrm{H}\alpha~A}=5.29\times10^{-15}$\ergpcmsqps and F$_{\mathrm{H}\alpha~B}=2.99\times10^{-15}$\ergpcmsqps. \label{fluxes} }
\end{table}

\subsubsection{[O {\small III}] emission}

A characteristic of the extended optical emission filaments in many BCGs is their low [O {\small III}] emission with respect to the hydrogen recombination lines (see for example \citealt{hatch2006}) and surprisingly high [Ne {\small III}] emission lines (see discussion of charge exchange in \citealt{ferland2009}). The ionisation potential of Ne$^{+}$ is greater than that of O$^{+}$ so if we observe [Ne {\small III}] lines we would expect to see those of [O {\small III}] as well. 

The total [O {\small III}]$\lambda$5007 emission in NGC 4696 is fairly strong with an [O {\small III}]$\lambda$5007/H$\beta~\lambda$4861 intensity ratio measured as 1.4 in the inner 27$\times$27 arcseconds (the field-of-view of our HRB grism). However, as discussed in Section \ref{fil_morph} the surface brightness of [O {\small III}] emission is steeply peaked and located near to the centre of radio emission. It is therefore probable that much of the [O {\small III}] emission is due to the central AGN. 

We investigate this possibility by choosing two regions; one coincident with the radio nucleus (region A) and one offset by $\sim$12 arcseconds, on the farthest edge of the inner filament from the nucleus (region B). Spectra are extracted in both regions from a 3 pixel (2'') radius circle, the regions are shown in Fig. \ref{inner_outer_circles} and the spectral regions containing the H$\beta$ line and [O {\small III}] emission are shown in Fig. \ref{oiii_hb_spectra}. The ratio of [O {\small III}]$\lambda$5007/H$\beta~\lambda$4861 clearly varies in the two regions with the central region having a ratio of $\sim$2 and the filament emission having a much lower [O {\small III}]$\lambda$5007/H$\beta~\lambda$4861 ratio of only $\sim$0.7. Ratios of the fluxes of the detected lines to H$\alpha$ for the two regions are shown in Table \ref{fluxes}.

\begin{figure*}
\centering
\includegraphics[width=0.7\textwidth]{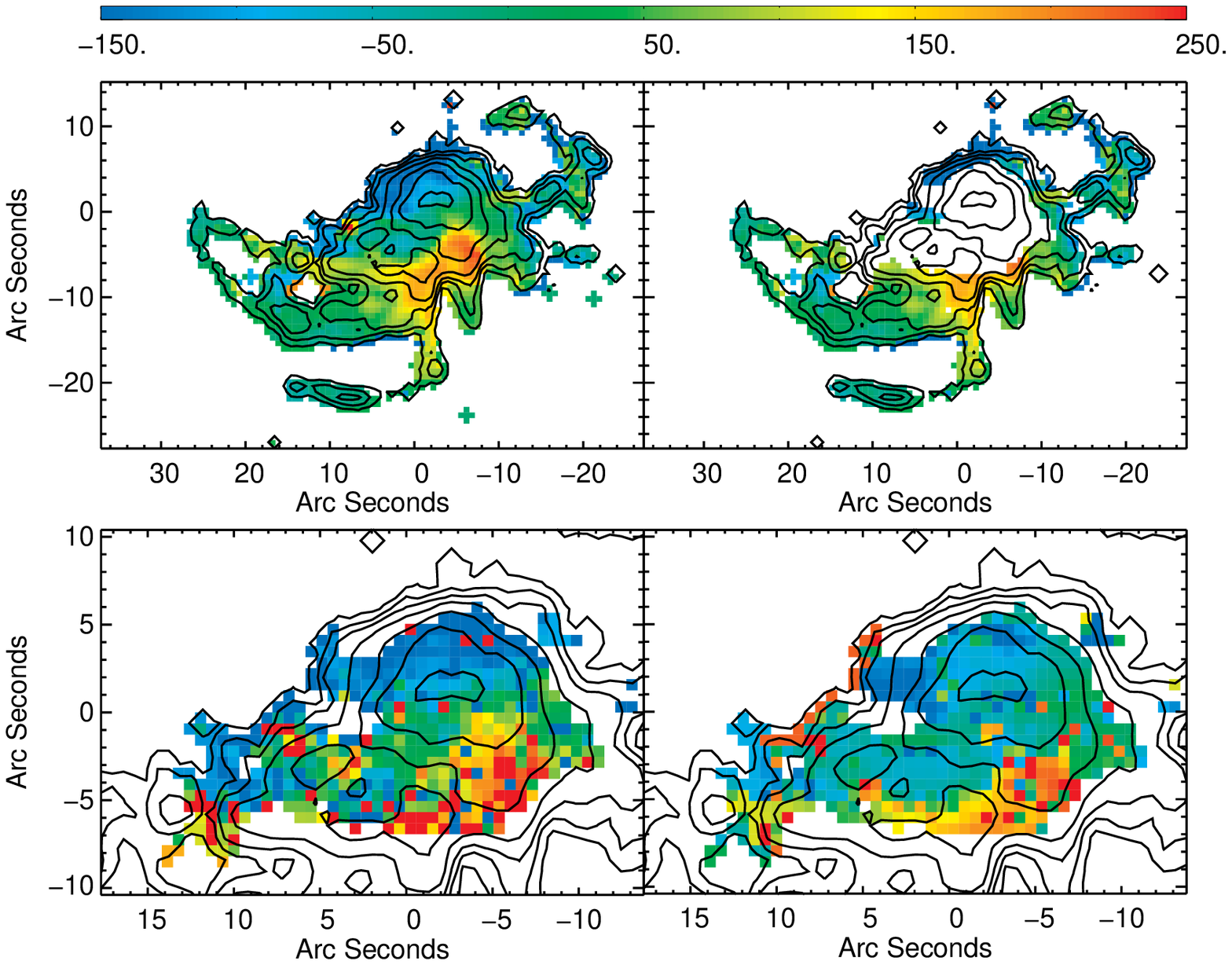}
\includegraphics[width=0.7\textwidth]{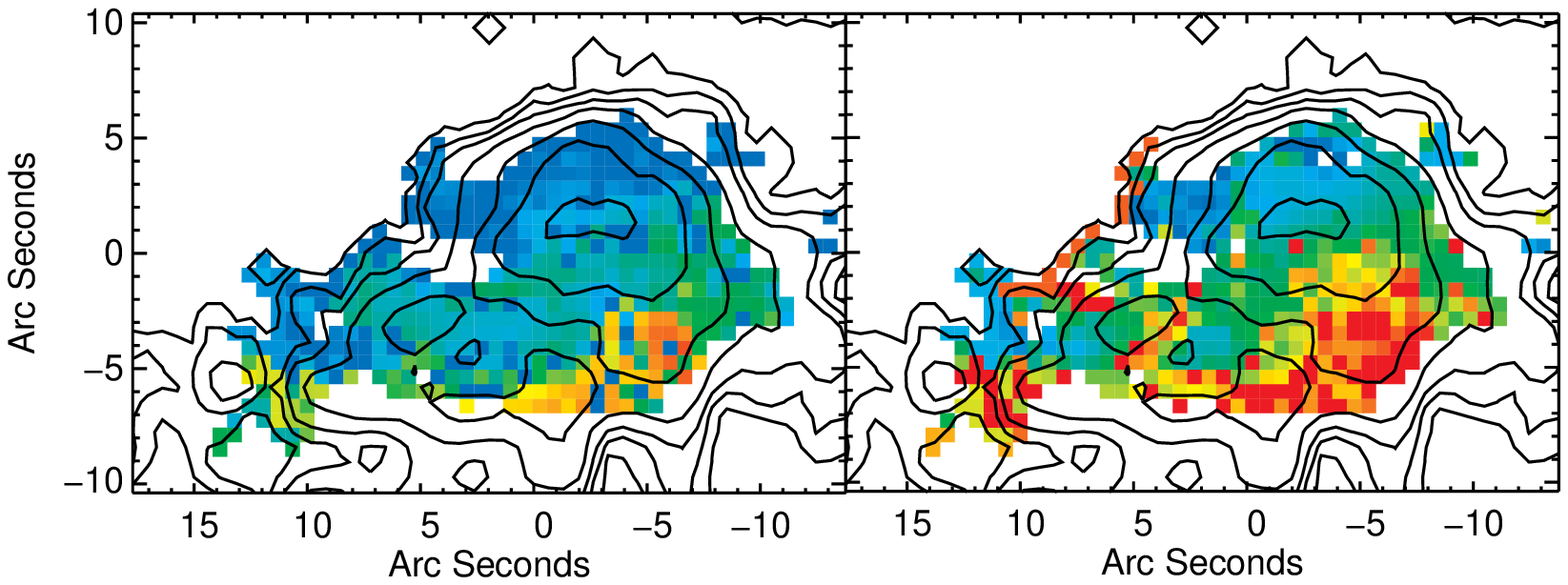}
\caption{Top: (Left) The line of sight velocity relative to the velocity of the galaxy, 3045~\kmps\ (z$=$0.010157),
from \protect \cite{postman1995}, if a single component fit is made over the whole field of view and
(right) the line of sight velocity shown only where one velocity component was
found to be sufficient. The figures shown are fit from the strong [N {\sc ii}]$\lambda$6583 emission line.
Middle: The same as the above for the two component fit. The left hand plot indicates 
the broader component and the right hand plot the narrower component. Bottom: The
blueshifted (left) and redshifted (right) velocity components.
The colour bar is in units of \kmps\ and the axis (0,0) position corresponds to RA and Dec of $12^{h}48^{m}49^{s}.28$, $-41^{\circ}$18'39.4'', coinciding with the core of the radio emission. \label{los_vel}}
\end{figure*}

\begin{figure*}
\centering
\includegraphics[width=0.7\textwidth]{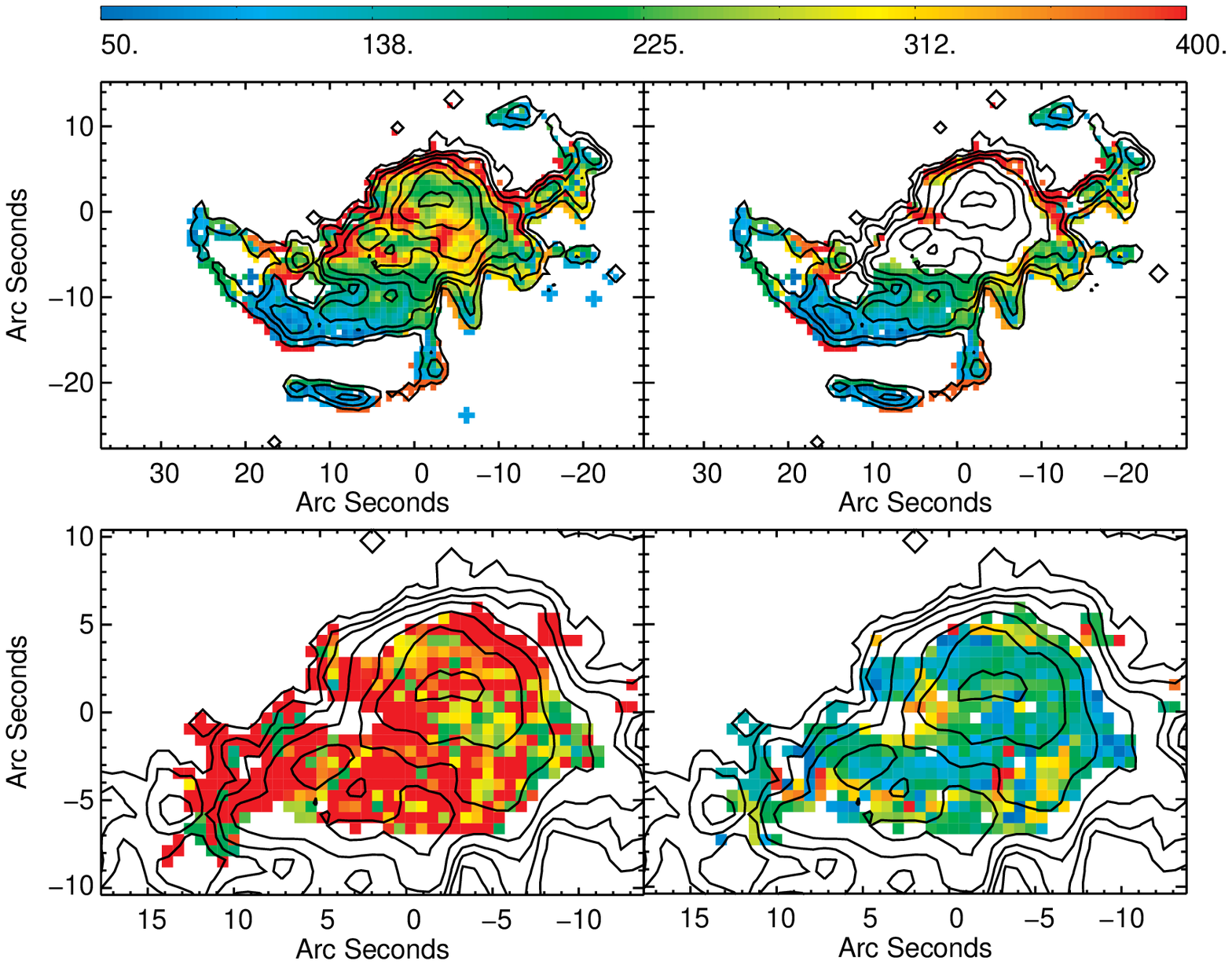}
\includegraphics[width=0.7\textwidth]{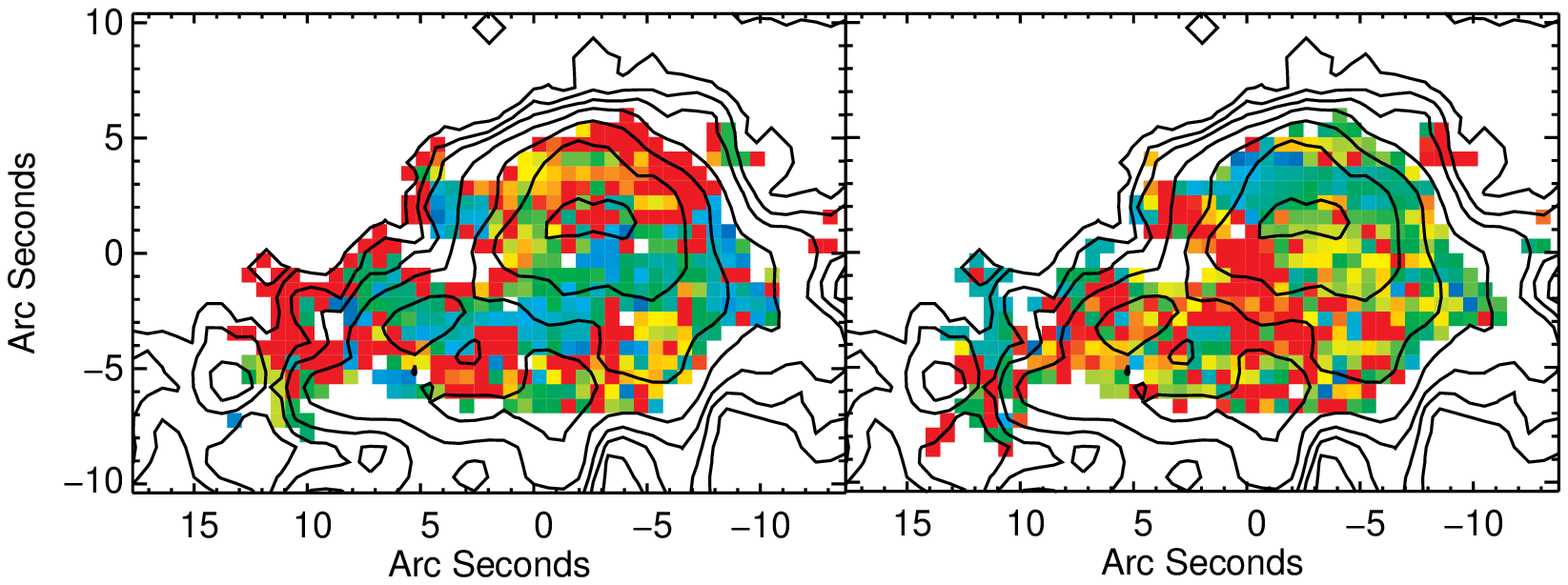}
\caption{Top: (Left) The velocity width at full width half maximum if a single
velocity component fit is assumed over the whole field of view and (right) the
velocity width where only one velocity component was found to be sufficient. The figures shown are fit from the strong [N {\sc ii}]$\lambda$6583 emission line.
Middle: The same as the above for the two component fit. The left hand plot indicates 
the broader component and the right hand plot the narrower component. Bottom: The
blueshifted (left) and redshifted (right) velocity components.
The color bar is in units of \kmps\ and has been corrected for the instrumental
width. The axis (0,0) position corresponds to RA and Dec of $12^{h}48^{m}49^{s}.28$, $-41^{\circ}$18'39.4'', coinciding with the core of the radio emission.
\label{vel_disp}}
\end{figure*}

Observations of the inner regions of NGC 1275 \citep{kent1979, sabra2000} have found [O {\small III}]$\lambda$5007/H$\beta~\lambda$4861 ratios of 0.6 in the inner 24 arcsec (8~\kpc), however the [O {\small III}]$\lambda$5007 emission in the outer filaments, beyond 10~\kpc, is not detected in the long slit spectra of \cite{hatch2006}. The authors are able to put an upper limit on the [O {\small III}]$\lambda$5007/H$\alpha~\lambda$6563 emission in a region 18~\kpc\ from the nucleus of $<$0.03, their limit on [O {\small III}]$\lambda$5007/H$\beta~\lambda$4861 is $<$0.13.

Using long slit spectra from the INT and FOS spectra \cite{johnstone1988} found that the [O {\small III}]$\lambda$5007/H$\beta~\lambda$4861 ratio decreases away from the core in both NGC 1275 and Abell 1795. In NGC 1275 their spectra extend to a radius of 8~\kpc\ where they find the ratio drops by a third from the central value to about 0.5. Similarly in Abell 1795 the ratio peaks at about 1 in the nucleus and decreases to $\sim$0.6 at the outer extent of the slit which correspond to a radius of 6~\kpc. Low [O {\small III}]$\lambda$5007/H$\beta~\lambda$4861 ratios have also been observed in the vast emission line nebula associated with the BCG of Abell 2597 \citep{voit1997}.

Shock models tend to predict higher [O {\small III}]$\lambda$5007/H$\beta~\lambda$4861 intensity ratios and larger fluxes of [O {\small III}]$\lambda$4363  than those seen in the outer extents of the emission line nebulae surrounding these objects \citep{sabra2000, voit1997}. We obtain only upper limits on the [O {\small III}]$\lambda$4363 emission across the field-of-view with the upper limit on the [O {\small III}]$\lambda$4363/H$\beta~\lambda$4861 intensity ratios of $\sim$0.08. Unfortunately we only acquired one pointing of HRB data in NGC 4696 so are unable to investigate the ratio in the very outer extent of the nebulae.

Our spectra do not extend far enough into the blue to observe the [Ne {\small III}]$\lambda$3869 optical line, however, \cite{farage2010} have detected this at the 3$\sigma$ level (see their Fig. 3). The infra-red line of [Ne {\small III}]$\lambda$15.56~$\mu$m has also been clearly detected using \textit{Spitzer} IRS SH spectra by \cite{johnstone2007}. Photoionisation models for the excitation of the emission line nebulae struggle to account for the anomalously low [O {\small III}] yet high [Ne {\small III}] emission unless coupled with high metallicities. \cite{sanders2006} find the oxygen and Ne abundances to peak at about solar towards the centre of the Centaurus cluster. However, the measured Ne abundance is highly dependent on the spectral model and varies between 0.5-2 between the SPEX and APEC models.

Particle ionisation models put forward by \cite{ferland2008, ferland2009} have so far been successful at reproducing the observed ratios due to the importance of charge-transfer. Using {\sc cloudy} simulations (last described by \citealt{ferland1998}) the authors find the charge-transfer recombination of O$^{+}$ and O$^{+2}$ is very fast compared to He$^{+}$, Ne$^{+}$ and Ne$^{+2}$ resulting in significant [Ne {\small II}] and [Ne {\small III}] emission compared with [O {\small III}].   

\begin{figure}
\centering
\includegraphics[width=0.45\textwidth]{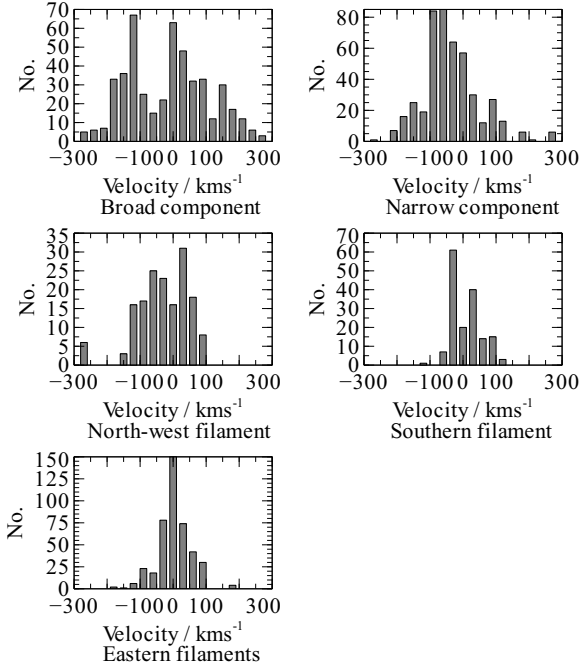}
\caption{The distribution of line-of-sight velocities in the filaments and central regions. \label{los_histograms}}
\end{figure}

\subsection{Velocity structure}
 
Maps of the kinematics of the optical emission line nebulae are shown in Figs. \ref{los_vel} and \ref{vel_disp}. The line-of-sight velocities are given with respect to the redshift of NGC 4696 determined to be $z=0.010157$ (3045 \kmps) by \cite{postman1995}. The velocity dispersions are presented as the FWHM velocities measured from the gaussian fits to the strongest emission lines in our spectra and have been corrected for the measured instrumental width, determined from fitting sky lines close to the observed H$\alpha$ emission line (FWHM 98~\kmps, $\sigma=42$\kmps). All emission lines strong enough to map were found to have essentially the same velocities, as seen in \cite{farage2010}, therefore we only show maps of [N {\small II}]$\lambda$6583.

\subsection{Central velocity structure}

The line-of-sight velocity of the central region of the galaxy changes smoothly by $\sim$400 \kmps\ across the north to south direction with the southern region being more redshifted, perhaps indicating a component of rotation of the nebula about the nucleus. This gradient in velocity structure is seen in both the broad and narrow velocity components but to different extents (see Fig. \ref{los_histograms}).

The broad component has a wider distribution of line-of-sight velocities, with the northern-most and southern-most velocities varying by $\sim$500 \kmps\ while in the narrow component the velocity changes by $\sim$250 \kmps\ from the northern-most to southern-most regions. This limited range in velocities is similar to those observed in the extended filaments seen to the north-west, south and east of the nucleus.

The broad velocity component has a FWHM of $\sim$300$-$400 \kmps\ while the narrow component has widths in the range 50$-$200 \kmps. The line widths of both components are much greater than the thermal widths of the hydrogen gas which is broadened by $\sim$10 \kmps\ at 10$^{4}$~K implying either the gas is broadened by shocks or by turbulence. 

\begin{figure}
\centering
\includegraphics[width=0.45\textwidth]{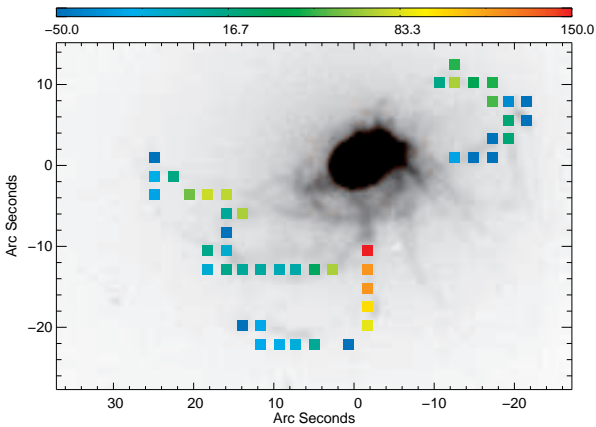}
\caption{The line-of-sight velocities in the outer filaments. The velocity zero-point is taken as 3045 \kmps. \label{outer_redshift}}
\end{figure}

\begin{figure}
\centering
\includegraphics[width=0.45\textwidth]{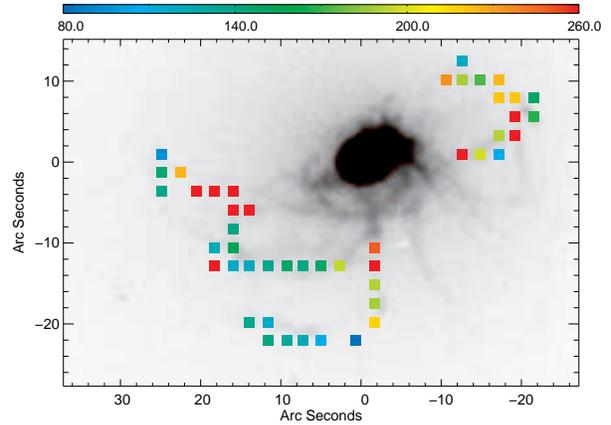}
\caption{The FWHM velocity widths in the outer filaments after correction for the instrumental width. \label{outer_velocity}}
\end{figure}

\subsection{Outer filaments}

The distribution of line-of-sight velocities of the outer filaments is less than the central region, with a range of $\sim$200 \kmps\ in most cases. We bin the outer emission in 3 by 3 arcsecond bins to examine the velocity structure more closely along the length of the filaments, the results are shown in Fig. \ref{outer_redshift}. The line-of-sight velocity zero-point is taken as 3045~\kmps.

\begin{figure*}
\centering
\includegraphics[width=0.9\textwidth]{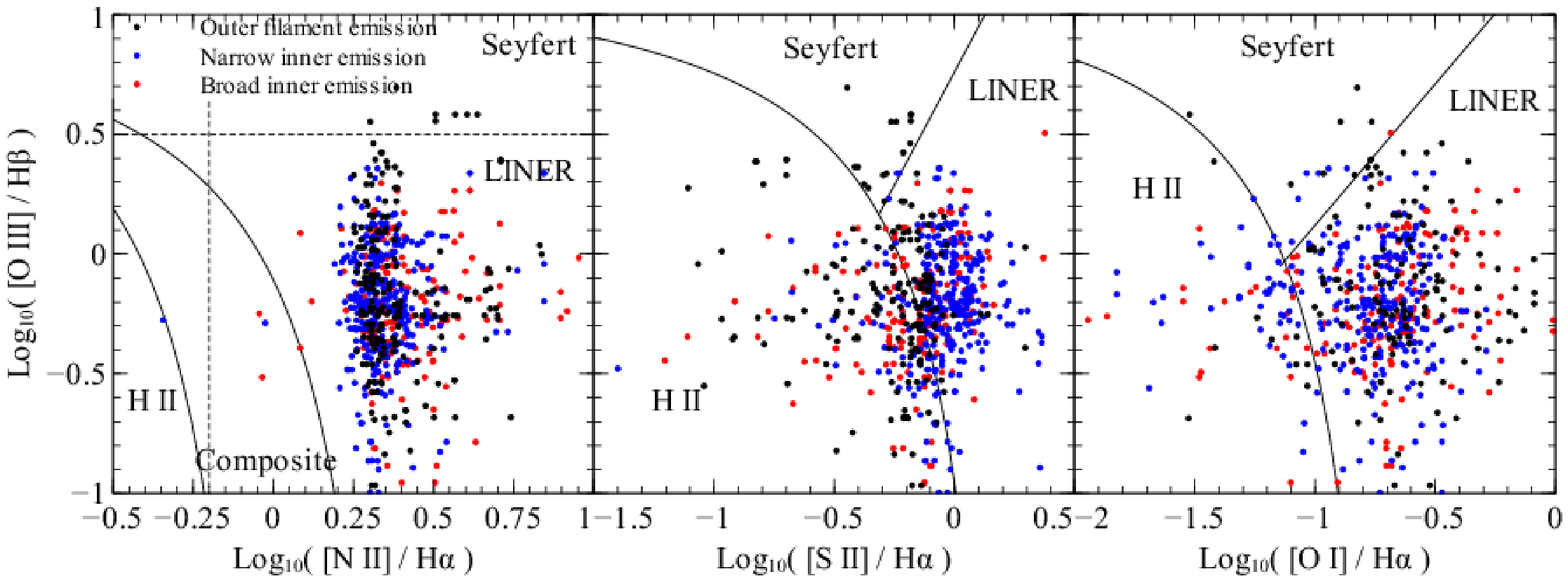}
\caption{Ionisation plots \protect \citep{baldwin1981} of the outer and inner regions in the central 27'' of NGC 4696. Top: [O {\small III}]$\lambda$5007/H$\beta~\lambda$4861 verses
[N {\small II}]$\lambda$6583/H$\alpha~\lambda$6563 emission. Middle:
[O {\small III}]$\lambda$5007/H$\beta~\lambda$4861 verses 
[O {\small I}]$\lambda$6300/H$\alpha~\lambda$6563 emission. Bottom:
[O {\small III}]$\lambda$5007/H$\beta~\lambda$4861 verses
[S {\small II}]$(\lambda6716+\lambda6731)$/H$\alpha~\lambda$6563 emission. Solid lines are from \protect \cite{kewley2006} and dashed lines from \protect \cite{osterbrock2006}.
\label{bpt}}
\end{figure*}

The eastern filament is the dominant structure curling round the south-east of the galaxy, coinciding closely with the outer dust lanes and soft X-ray filaments. The H$\alpha$ imaging of \cite{crawford2005} shows this filament is in fact made of several thinner threads such as those resolved in NGC 1275 \citep{fabian2008}. The filament extends over 30'' (6~\kpc) from the nucleus but exhibits a very narrow distribution in line-of-sight velocities peaking at the same velocity as the galaxy (see Fig. \ref{los_histograms}). The low velocities suggest we are vewing this filament close to the plane of the sky. Both the eastern filament and the north-west filament appear to curve round the radio bubble. The north-west filament has a larger variation in line-of-sight velocities, though the variation is still smooth, with the innermost part more blueshifted and the outermost part more redshifted.

The southern filament exhibits a divide in line-of-sight velocity along the filaments length such as was found by \cite{hatch2006} in the northern filament of NGC 1275. The northern most part, closest to the nucleus, displays a redshift of $\sim$140 \kmps\ whereas the region farthest from the nucleus displays a blueshift of $\sim-$50 \kmps. The regions closest and farthest from the nucleus along this filament may thus be moving in opposite directions with the filament either collapsing in or stretching out, however the filament is clearly curved and this scenario is complicated by projection effects.

The velocity widths of the emission lines in the outer filaments are fairly low, ranging from 50$-$250 \kmps\ with the more northern filaments having a slightly larger velocity width than the southern ones. The range of velocity widths measured in these filaments is similar to those seen in the Perseus cluster \citep{hatch2006}.

\subsection{Implications for filament origin}

Kinematically the central narrow emission line component bears more similarity to the outer filaments than the broad component having both a lower velocity dispersion and a narrower distribution of line-of-sight velocities. This supports the interpretation of the narrow component being a separate filament farther from the nucleus, seen in projection with the main central broad component. This would suggest the filamentary system is not an intrinsically one-sided spiral structure and it is unclear how a merger origin could account for the diversity in morphology and kinematics of these filaments.

If the filaments had an inflow origin, as predicted by early cooling flow models \citep{fabian1984}, we would expect that there would be a negative radial gradient in the widths of the emission lines due to both greater acceleration towards the centre and line-of-sight effects \citep{heckman1989}. This is not observed in any of the outer filaments with the velocity dispersion being largely unvarying with the exception of the north-west filament where the velocity dispersion appears erratic along the filament. This filament exhibits knots in the H$\alpha$ image, just north-east of the bend in the filament and again at the end. Another fainter filament is also present in the H$\alpha$ image, which crosses or joins the north-west filament at this bend where the measured velocity width is largest (see Fig. \ref{outer_velocity}). It is therefore likely that the measured velocity width in this region is large and erratic due to two or more filaments overlapping, in the line-of-sight, with slightly different velocities and resulting in a broad, strong peak of emission.

\cite{crawford2005} suggest the filamentary system is the consequence of buoyantly rising radio bubbles which draw filaments of cool and cold gas and dust up beneath them as has also been seen in other brightest cluster galaxies \citep{bohringer1995, churazov2001, reynolds2005, fabian2003, hatch2006}. This is supported by a depression in the X-ray intensity and thermal pressure maps to the north-east of the nucleus, just beyond the optical and soft X-ray filaments (see their Fig. 5). A similar depression is seen in the south-west and low frequency (330 MHz) radio emission also extends towards both these structures. The kinematics of the filaments are consistent with this picture with smoothly varying line-of-sight velocities and low velocity dispersions which show little variation across the whole length of some filaments.

\subsection{Emission line ratios}

\begin{figure}
\centering
\includegraphics[width=0.45\textwidth]{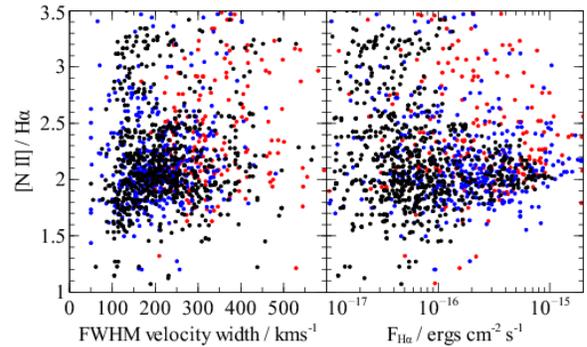}
\caption{[N {\small II}]$\lambda$6583/H$\alpha~\lambda$6563 ratio plotted against the FWHM velocity with of each pixel and against the H$\alpha$ flux. Only pixels where both [N {\small II}]$\lambda$6583 and H$\alpha~\lambda$6563 were detected at the 4 $\sigma$ level are plotted. Colour codeing same as in Fig. \ref{bpt}. \label{nii_ha_total}}
\end{figure}

The emission line ratios in the optical nebulosity are important tracers of the excitation processes dominating their behaviour. \cite{heckman1989} suggested there was a bimodal population of BCGs based on their [N {\sc II}]/H$\alpha$ ratios and H$\alpha$ luminosity. However, using a much larger cluster sample \cite{crawford1999} have shown that this is in fact a continuous spectrum with all objects sharing the characteristic low-ionisation emission line spectrum. 

In a study by \cite{wilman2006} of 4 high H$\alpha$ luminosity BCGs all situated in cool core clusters and with the same LINER emission properties, the authors found no evidence for variations in line ratios as a function of either position or H$\alpha$ luminosity across the galaxies. They interpret this as evidence for both a single ionisation source for the line emission across each galaxy and a single dominant excitation mechanism in all of their galaxies. \cite{hatch2007} found, in a similar study of optical nebulosities surrounding 6 BCGs spanning a range of optical, X-ray and radio properties, that the ionisation state of the optical nebulosities were not uniform and so concluded the converse; that a single scenario cannot be responsible for the optical line emission in their sample. A recent study by \cite{edwards2009} of a further 9 clusters, including in their sample both cool and non-cool core clusters, found diverse emission line ratios and morphologies. This supports the finding by \cite{hatch2007} that there is no one clear consistent mechanism to explain the origin of the line emission or to separate BCGs in cool and non-cool core clusters. Despite this there are some clear similarities in that a relatively hard ionising source is needed and that the systems are always well described by a LINER-like emission line spectrum.

\cite{hatch2006} studied in detail the spectral features in NGC 1275, a similar object in the centre of the Perseus cluster, and found radial variations in the [N {\sc ii}]/H$\alpha$ emission line ratio along the filaments with the inner regions having higher [N {\sc ii}]/H$\alpha$ ratios. This could be due to an extra energy source closer in causing more heating per hydrogen ionisation or due to metallicity variations in the filaments. \cite{sarzi2006} also found variations in ratios of [O {\sc ii}]/H$\beta$ along the length of the filaments in M87, the central galaxy in the Virgo cluster. 

Maps of our emission line ratios, binned using the contour binning algorithm of \cite{sanders2006c} are shown in Fig. \ref{ratios}. The top panel showing the single gaussian fits, the middle showing the broad central velocity component and the bottom panel in each case shows the narrow component. Only pixels where the line emission is detected at greater than 4$\sigma$ are shown. A ionisation plot \citep{baldwin1981} of the same pixels is shown in Fig. \ref{bpt}. Black points indicate the pixels where one emission line component was found sufficient to model the emission lines, red points correspond to the broad lines and blue to the narrow. The [O {\sc iii}]]/H$\beta$ ratios cannot be traced across the whole length of the outer filaments as our HRB data only cover the central 27 arcseconds, as such we cannot plot the most extended parts of the outer filaments on Fig. \ref{bpt}.  

The scatter in Fig. \ref{bpt} is large, however NGC 4696 clearly exhibits LINER-like emission in both the central and extended regions. The broad inner component has in general a higher [N {\sc II}]/H$\alpha$ ratio than the narrow or outer regions, which are virtually indistinguishable from each other. If the broad component is from more centrally located emission this is consistent with the observation of \cite{hatch2006} in NGC 1275 that [N {\sc ii}]/H$\alpha$ ratios are in general higher towards the centre of the galaxy. If the filaments are shock excited, we should see a gradient in emission line ratios with velocity dispersion of the gas. Fig. \ref{nii_ha_total} shows this ratio for the same three regions as in Fig. \ref{bpt}. No correlation is seen in [N {\sc ii}]/H$\alpha$ ratio with velocity width with most outer filaments (black points) having FWHM velocity widths of 100-200 \kmps\ but ratios which vary from 1.8-3.3. We now examine the variation in emission line ratios across the outer filaments in more detail by binning the emission in 3'' by 3'' regions across the field-of-view (see Fig. \ref{nii_ha_outer}).

\subsection{Outer filaments}

\begin{figure}
\centering
\includegraphics[width=0.45\textwidth]{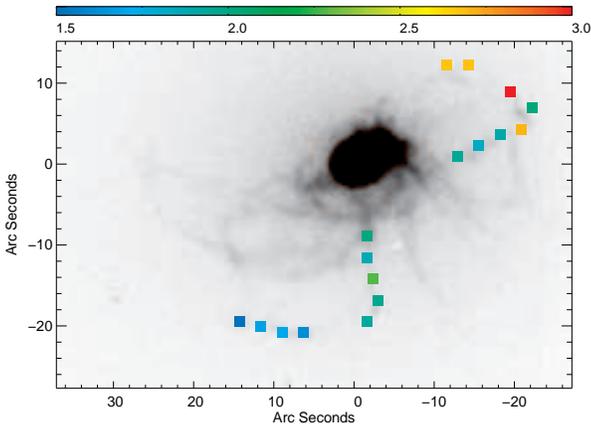}
\caption{[N {\small II}]$\lambda$6583/H$\alpha~\lambda$6563 emission ratios in the outer north-west and southern filaments. The emission is binned in 3''$\times$3'' regions across these filaments. Only regions where both emission lines were detected at greater than 4 sigma are shown. \label{nii_ha_outer}}
\end{figure}

\begin{figure}
\centering
\includegraphics[width=0.45\textwidth]{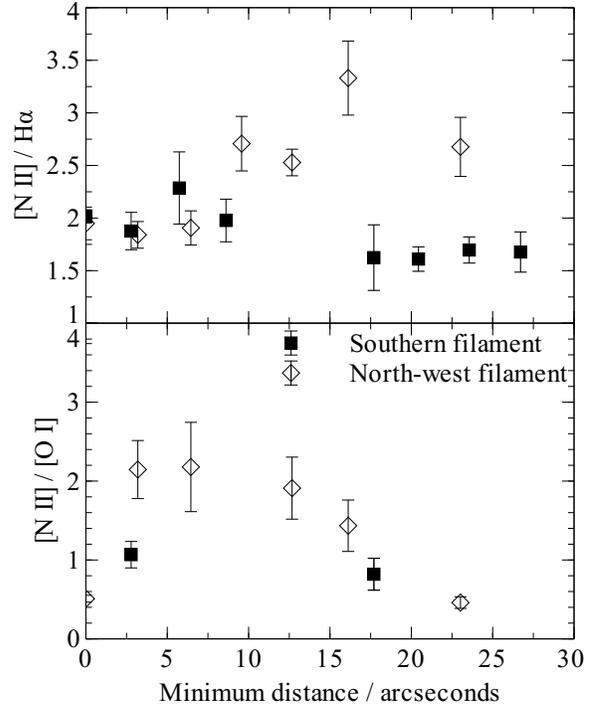}
\caption{Variation in emission line ratios as a function of projected distance along the north-west and southern filaments. Regions where spectra were extracted are shown in Fig. \ref{nii_ha_outer}. \label{ratios_outer}}
\end{figure}

The north-west and southern filaments show a difference in the [N {\small II}]$\lambda$6583/H$\alpha~\lambda$6563 emission ratios across their projected length with the southern filament decreasing and the north-western filament increasing in [N {\small II}]$\lambda$6583/H$\alpha$ ratio (Fig. \ref{nii_ha_outer}and Fig. \ref{ratios_outer}). The innermost regions of both filaments have a ratio $\sim$2, the average ratio of the galaxy. This difference in variation is in contrast to that seen in the filaments of NGC 1275 by \cite{hatch2006} where there is a clear radial trend in [N {\small II}]$\lambda$6583/H$\alpha$ intensity ratio seen in all long slits, with the ratio decreasing with projected distance from the nucleus.

\begin{figure}
\centering
\includegraphics[width=0.45\textwidth]{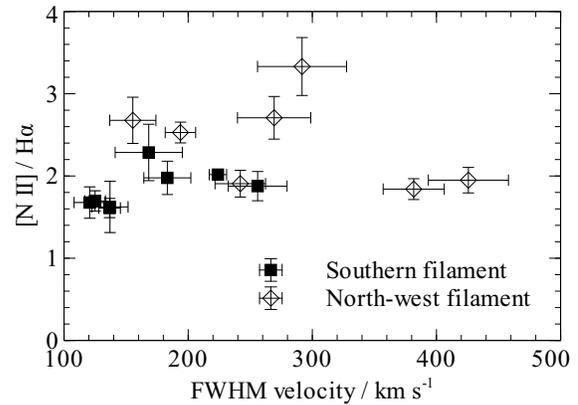}
\caption{[N {\small II}]$\lambda$6583/H$\alpha~\lambda$6563 emission ratio as a function of FWHM velocity width in the north-west and southern filament region. \label{ratio_velocity_outer}}
\end{figure}

A higher [N {\small II}]$\lambda$6583/H$\alpha$ intensity ratio indicates that there is more heating occurring in the gas per hydrogen ionisation. The difference in emission line ratios in these two filaments could be due to the proximity of the filaments to the expanding radio bubble. The arm of the north-western filament appears to encircle the radio bubble while the southern filament appears spatially less connected with the current radio emission.

The [N {\small II}]$\lambda$6583/H$\alpha$ intensity ratio is also sensitive to variations in the metallicity of the gas. X-ray observations of the Centaurus cluster \citep{sanders2008b} have shown there is a high abundance of nitrogen emission in the central regions. Oxygen is an efficient coolent and high oxygen abundances in the gas would act to decrease the gas temperature and so reduce the observed [N {\small II}]$\lambda$6583/H$\alpha$ ratio, however the nitrogen/oxygen ratio in the hot gas is also high. The bottom panel of Fig. \ref{ratios_outer} shows the trend in [N {\small II}]$\lambda$6583/[O {\small I}]$\lambda$6300 emission line ratios in these same filaments. There is weak evidence for a decreasing trend in [N {\small II}]$\lambda$6583/[O {\small I}]$\lambda$6300 ratio in these filaments.

The variation of [N {\small II}]$\lambda$6583/H$\alpha$ emission line ratio with FWHM velocity width of the gas, in these two outer filaments, is shown in Fig. \ref{ratio_velocity_outer}. There is no correlation between the measured ratios and the width of the lines in either of the southern or north-western filaments. The majority of the emission has velocity widths less than 250~\kmps\ and a emission line ratio of $\sim$1.8-2.5. The two very broad regions in the north-west filament correspond to bright knots of emission which could be formed by the intersection of two or more individual filaments which cannot be resolved in our data. 

\section{Conclusions}

The origin and excitation of the complex extended emission line nebula in BCGs and more specifically in NGC 4696 has been a matter of some controversy. Fig. \ref{inner_and_dust} shows the close correspondence between the dust absorption and the [N {\sc ii}] emission tracing the 10$^{4}$~K emission-line nebula in NGC 4696. Virtually all extended filaments are spatially coincident with dust emission. \cite{sparks1989} do not find any deviations in the dust extinction behaviour with wavelength with respect to dust within our Galaxy, similar results have been found in other BCGs. The authors suggest that this favours a merger origin for the emission line filaments over the formation of the filament system straight from the hot gas. 

There is however, a third possibility; that the gas originated due to dusty star formation in AGB atmospheres in the brightest cluster galaxy and has been drawn out, rather than falling in, by buoyantly rising radio bubbles inflated by the AGN \citep{crawford2005}. This scenario is supported by \chandra\ X-ray observations that reveal pressure minima towards the north-east at the end of the soft X-ray filaments and another minima towards the south-west. These minima coincide with the outer edge of the 330 MHz radio emission and appear to be surrounded by regions of higher metallicity X-ray gas \citep{fabian2005b}.

Our deep optical IFU data allow us to explore in detail the morphology, kinematics and ionisation properties of the optical nebulosity surrounding NGC 4696. Our observations show:
\begin{enumerate}
 \item there are at least two velocity components to the emission line gas in the inner regions of the galaxy. We interpret these as `broader' and `narrower' velocity width components,
 \item these components have a different morphology, with the narrow component having a less smooth morphology and a striking bar of bright emission while the broad component exhibits two peaks in emission, one peak coincident with the radio core and the other 4 arcsec to the west (see Fig. \ref{morph}),
 \item the peak in [O {\sc iii}] emission is spatially coincident with the nucleus and is offset from the brightest peak in [N {\sc ii}] emission (see Fig. \ref{morph_oiii}),
 \item the kinematics and emission line spectra of the extended filaments are similar to filaments surrounding other BCGs,
 \item we do not find any evidence to suggest there is a correlation between the velocity width and [N {\sc ii}]/H$\alpha$ ratio in the filaments of NGC 4696,
 \item there is very little [O {\sc iii}] emission in the outer filaments as observed in similar systems.
\end{enumerate}

The central 12 arcseconds radius in NGC 4696 contains at least two components to the emission line gas, we interpret these as a broad velocity component and a second narrower component. The reddening and kinematical properties of this second component lead us to suggest it is a filament with properties much like the other extended filaments, that is located behind the main galaxy and thus more obscured by the intervening dust lanes. The morphology of the central region was previously thought to be a simple one-sided spiral structure perhaps caused by the tail of an infalling galaxy \citep{sparks1989, farage2010} however the filament system is clearly complex and it is not immediately obvious how a merger could account for the diversity in morphology of the filaments.

We find smoothly varying line-of-sight velocities in the outer filaments and low velocity widths (FWHM 50-250~\kmps), with little variation along the length of many filaments, similar to those found in NGC 1275, the brightest cluster galaxy in the Perseus cluster \citep{hatch2006}. This is consistent with an origin for the filamentary system being slowly drawn up under the rising radio bubbles. 

The spectrum of NGC 4696 exhibits the same strong low ionisation emission lines seen in many other extended emission line systems surrounding BCGs. These systems are not all consistent with a merger origin. \cite{johnstone2007} detected strong [Ne {\sc iii}]$\lambda$15.56$\mu$m emission in their \textit{Spitzer} IRS SH spectroscopy of NGC 4696 and NGC 1275. Optical [Ne {\sc iii}] line emission has also been seen in both NGC 1275 \citep{hatch2006} and weakly in NGC 4696 \citep{farage2010}. However, [O {\sc iii}] emission tends to be low in the extended emission line systems in BCGs (e.g. \citealt{donahue2000}). \cite{johnstone2007} have shown that stellar photoionisation models struggle to account for the high [Ne {\sc iii}] and low [O {\sc iii}] emission unless coupled with high metallicities. X-ray metallicities of typically solar for oxygen and neon \citep{sanders2006} are found in the centre of the Centaurus cluster and there is little sign of young stellar populations in the filaments.

The [O {\sc iii}]$\lambda$5007/H$\beta$ emission in NGC 4696 drops dramatically from the nucleus to $\sim$12 arcsec, indicating the [O {\sc iii}] emission is low in the filaments and the majority of emission in the inner regions is associated with the nucleus. A similar trend is seen in many other BCGs (e.g. \citealt{johnstone1988}). Our HRB data does not extend to beyond the inner filament so we cannot measure the [O {\sc iii}] emission in the farthest filaments from the nucleus. The ratio of our detected [O {\sc iii}] lines is small compared to shock models and the upper limit of the weak [O {\sc iii}]$\lambda$4364 line indicates a [O {\sc iii}]$\lambda$4364/H$\beta$ ratio of 5 per cent; lower than the predicted value from the best velocity shock model of \cite{farage2010}. 

\cite{ferland2008, ferland2009} have shown that the characteristic LINER spectrum with anomalously low [O {\sc iii}] and high [Ne {\sc ii}] and [Ne {\sc iii}] emission, compared to photoionisation models, can be reproduced with their particle heating model. With the exception of [N {\sc ii}]$\lambda$6583, which is under predicted, all our optical emission lines detected in the outer filaments in NCG 4696 are within a factor of 2 or less of their predicted spectrum. The predicted spectrum of \cite{ferland2009}, uses a gas-phase abundance of nitrogen relative to hydrogen of 6.9$\times10^{-5}$ which corresponds to $\sim$0.5Z$_{\odot}$ using abundances from \cite{anders1989}. \cite{sanders2008b} find the nitrogen abundance in Centaurus to be $\sim$4 times solar in the central regions, a factor of 8 times that used in the predicted particle heating spectrum. The predicted emission line ratios do not in general behave linearly with changes in metallicity of the gas due to the thermostat effect, where increasing the abundance of a coolant lowers the kinetic temperature to conserve
energy \citep{osterbrock2006}. However, the emissivities of [N {\sc ii}] and H$\alpha$ are predicted to be relatively constant over a broad range of temperatures ($10^{3.8}-10^{4.4}$~K) within the thermally stable regime (see Fig. 17 of \citealt{ferland2009}) so an increase in the nitrogen abundance by 8 would produce an increase in the [N {\sc ii}]/H$\alpha$ ratio. Note though that this is not necessarily the case for the [N {\sc i}] emission which does not exhibit such a flat emissivity profile. Future papers will explore the effect of varying metallicity on the particle heating model ratios.

\section{Acknowledgements}
REAC acknowledges STFC for financial support. ACF thanks the Royal Society. 
REAC would also like to thank Ryan Cooke, Bob Carswell and Paul Hewett for 
interesting and enlightening discussions.

This research has made use of the NASA/IPAC Extragalactic Database (NED) 
which is operated by the Jet Propulsion Laboratory, California Institute of 
Technology, under contract with the National Aeronautics and Space 
Administration.

\bibliographystyle{mnras}
\bibliography{/home/bcanning/Documents/latex_common/mnras_template}

\appendix

\section[]{Data reduction plots}
\label{appendixa}

\subsection{Sky absorption and emission corrections}

The VIMOS IFU does not have dedicated sky fibres and unlike slit spectroscopy an 
individual spectrum on a fibre does not have both regions with pure sky emission 
and regions with object and sky emission.
Different fibre spectra are described by different profile shape parameters 
such as the FWHM and skewness. In the wavelength direction this results in the 
the spectral lines having different shapes and thus affects the quality of the 
sky subtraction. An average sky spectrum formed by combining spectra with different 
line profiles can result in the presence of s-shape residuals when subtracted
from the data. VIPGI sky 
subtraction attempts to correct for these line profiles by grouping the 
spectra according to a user defined sky line. An average sky for each spectrum 
grouping is then calculated and subtracted. The groupings are done on a 
statistical basis thus this sky subtraction method employed by VIPGI is 
optimised for deep survey observations where the field is devoid of extended 
objects. It is not ideal for observing large galaxies which dominate the field 
of view. For this reason we performed the sky subtraction using
specific fibres shown to be lacking in emission lines associated with NGC 4696.

\begin{figure*}
\centering
\includegraphics[width=0.6\textwidth]{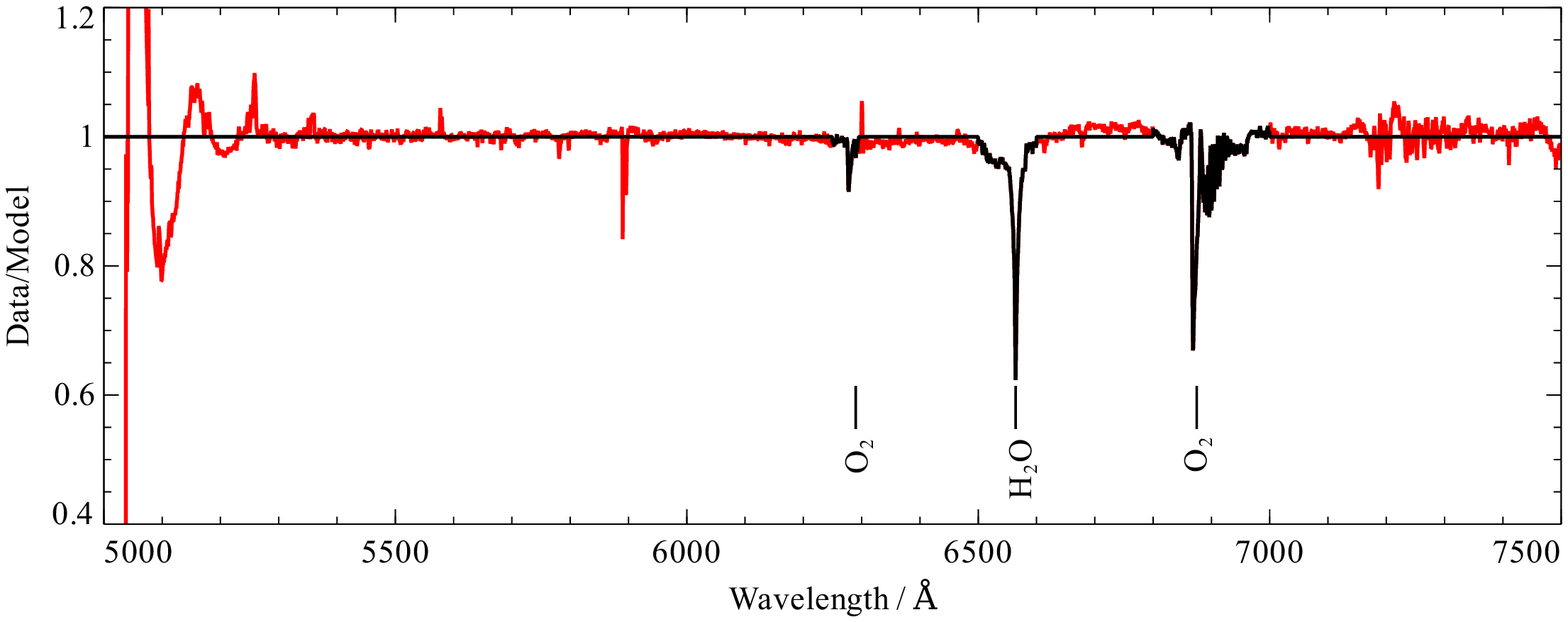}
\caption{Telluric absorption correction for Q1 using four standard star observations. We only 
correct for atmospheric absorption between $5500\,\mathrm{\AA}$ and $7000\,\mathrm{\AA}$. The 
absorption seen at $\sim 5900\,\mathrm{\AA}$ is likely NaD absorption from our Galaxy and
as such has not been corrected for. The red line shows the median observed spectrum of the four standard stars divided by their intrinsic spectrum. The black line is the teluric correction we apply to our data. \label{telluric}}
\end{figure*}

Telluric absorption feature corrections for the $\mathrm{O_{2}}$ and $\mathrm{H_{2}O}$ absorption 
in the $6000-7000\,\mathrm{\AA}$ band was determined from 4 observations of standard stars. Standard stars were observed at 
the beginning and the end of the night.
The stars chosen are CD-32-9927, LTT-7379, Hiltner-600 and LTT-2415.
The standard spectra $G(\lambda)$ were reduced using VIPGI, 
as with the science observations. These spectra were divided by the ESO spectra $G_{0}(\lambda)$
of the respective standards, corrected for atmospheric absorption. These observations were then 
averaged for each VIMOS IFU quadrant and the area around the absorption lines were isolated. Fig. \ref{telluric} 
shows the quadrant one spectra of each of the four standard stars and the averaged normalised 
spectrum.

Ignoring the curvature of the Earth we recover the object spectrum as seen without the atmospheric 
absorption ($F_{0}$),
\begin{equation}
 F_{0}(\lambda)=F(\lambda)\left[\frac{G_{0}(\lambda)}{G(\lambda)}\right]^{\frac{X_{\mathrm{object}}}{X_{\mathrm{standard}}}},
\end{equation}
for each 15 minute exposure separately so as to correct for airmass differences between the 
standard star ($X_{\mathrm{standard}}$) and the object ($X_{\mathrm{object}}$).

\subsection{Propagation of uncertainties}

\begin{figure}
\centering
\includegraphics[width=0.45\textwidth]{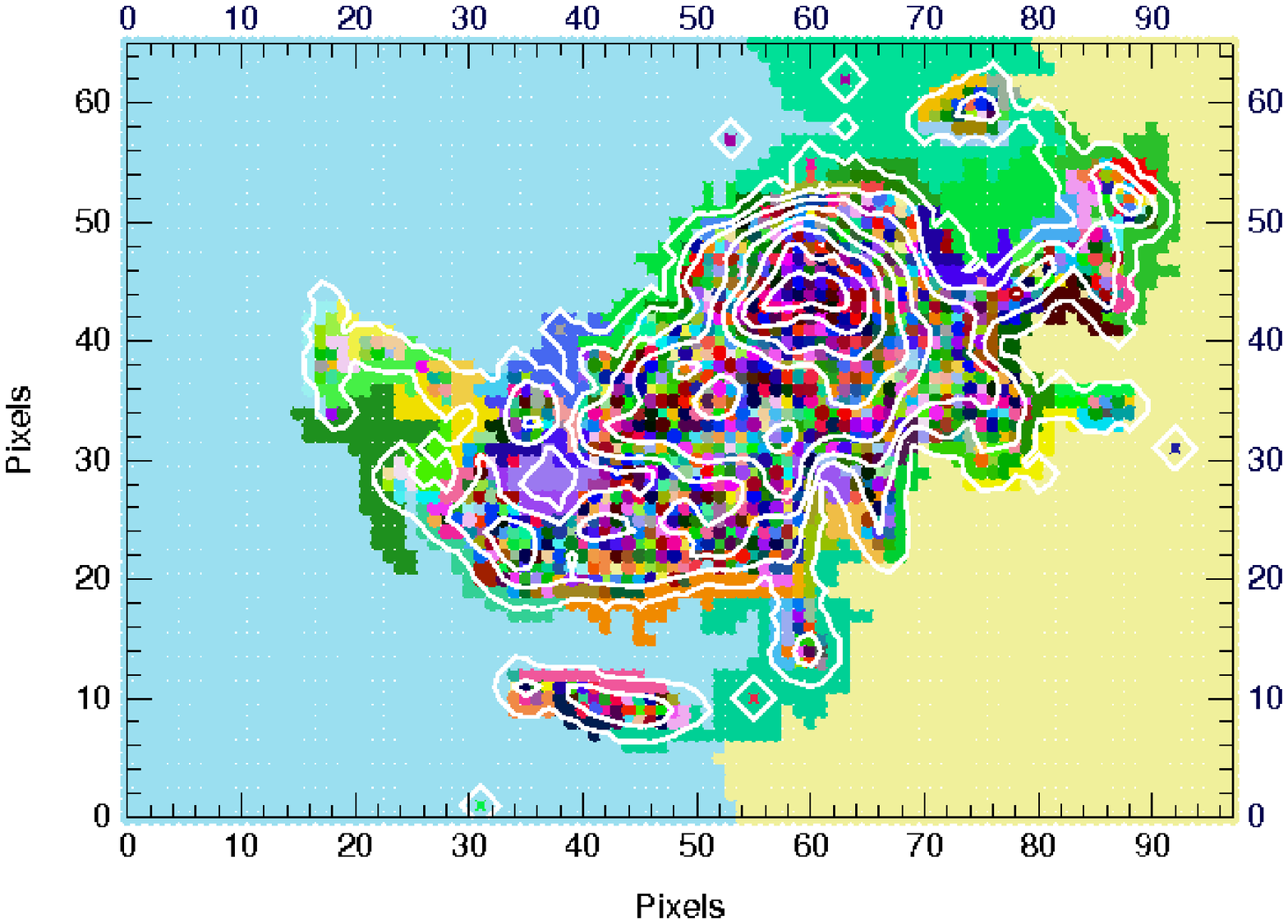}
\includegraphics[width=0.45\textwidth]{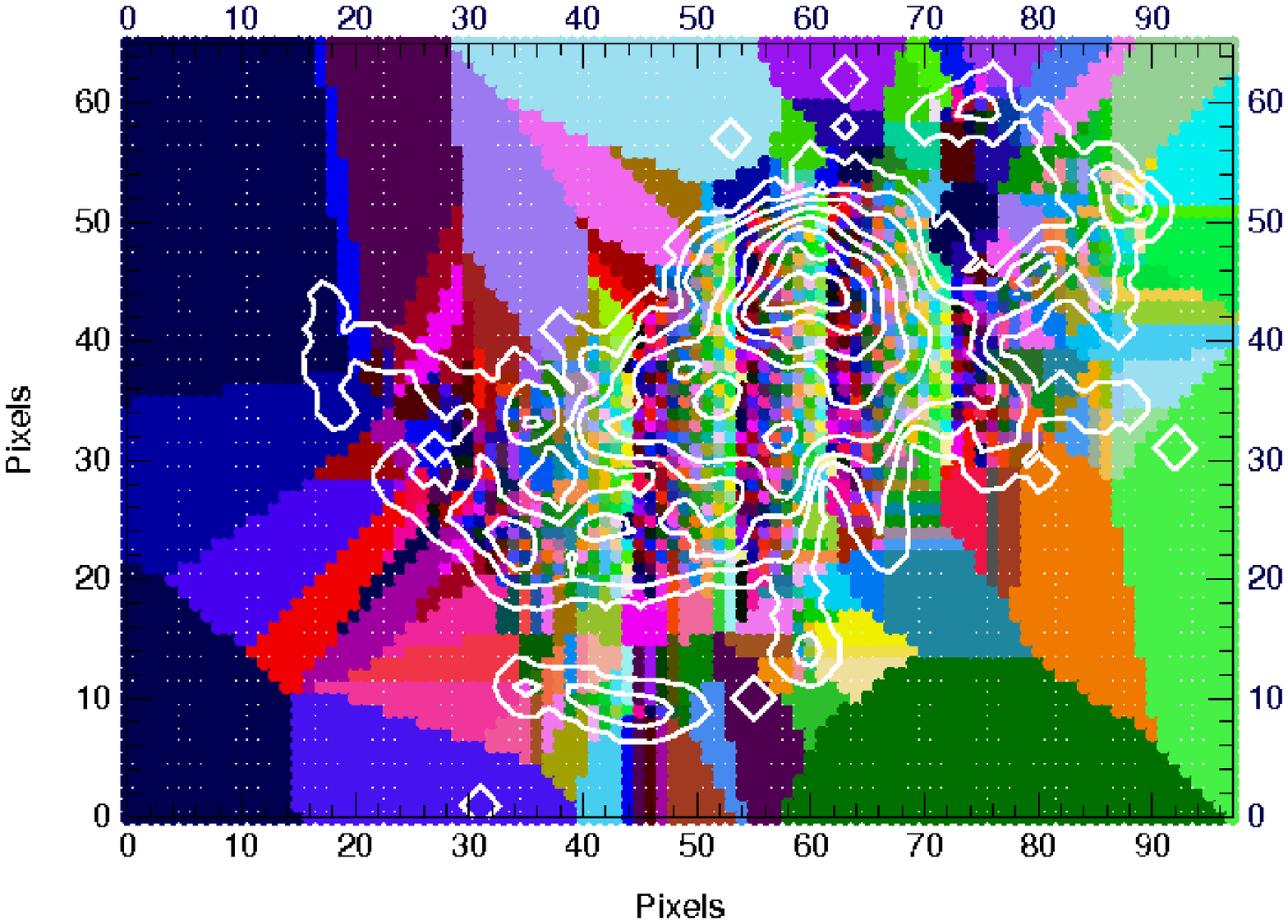}
\caption{Top: The contour binning algorithm of \protect \citep{sanders2006c} 
applied to a 2D brightness map of [N {\small II}]$\lambda$6583 emission in
NGC 4696. Bottom: The same using the Veronoi tessellation binning algorithm of
\protect \cite{cappellari2003}. Both maps show the data binned to a signal-to-noise
of 10. \label{binning}}
\end{figure}

Our HRO and HRB spectra overlap between $5000-6000~\mathrm{\AA}$. In order to incorporate
all our data we split the cubes into three wavelength regions. The regions chosen are
$4000-5100~\mathrm{\AA}$, $5100-5800~\mathrm{\AA}$ and $5800-7500~\mathrm{\AA}$. The 
two HR grisms used have a slightly different spectral resolution (see Table \ref{data_table})
with the HRO being marginally larger than that of the HRB grism. In order to amalgamate
both data sets it is necessary to re-bin the HRO spectra. If we are re-binning original bins 
$i$ into bins $j$ where some fraction $w_{ij}$ of $i$ falls into the $j$th bin we determine
the flux $F_{j}$ and error $S_{j}$ in the $j$th bin by,

\begin{equation}
 F_{j}\pm S_{j}=\sum_{i}w_{ij}f_{i}\pm \sqrt{\sum_{i}w_{ij}e_{i}^{2}},
 \label{sum}
\end{equation}

where $f_{i}$ is the flux in original bin $i$ and $e_{i}^{2}$ is the variance in bin
$i$. For pixel to pixel fluctuations the error $E_{j}$ is given by 
$E_{j}=\sqrt{\sum_{i}w_{ij}^{2}e_{i}^{2}}$.
The HRO and HRB spectra are then summed and the total flux $\mathcal{F}_{j}$ in the
re-binned grid is given by (Bob Carswell, private communication\footnote[2]{ftp://ftp.ast.cam.ac.uk/pub/rfc/vpfit9.5.pdf}),

\begin{equation}
 \mathcal{F}_{j}=\frac{\sum_{k}\frac{F_{j}^{k}}{(S_{j}^{2})^{2}}}{\sum_{k}\frac{1}{(S_{j}^{k})^{2}}},
 \label{totalflux}
\end{equation}
 
where $k$ is the number of spectra summed together. The estimate of the fluctuations 
in the spectra are then given by,

\begin{equation}
 \mathcal{E}_{j}^{2}=\frac{\sum_{k}(\frac{E_{j}^{k}}{(S_{j}^{k})^{2}})^{2}}{(\sum_{k}\frac{1}{(S_{j}^{k})^{2}})^{2}},
 \label{error}
\end{equation}

this forms the error spectrum used when fitting gaussian models to the emission lines
and minimising the resulting $\chi^{2}$.

\subsection{Spatial binning}

To improve the signal to noise in the spectra we investigate the use of two spatial binning algorithms; that of \cite{sanders2006c}, which bins the data based on surface brightness and that of \cite{cappellari2003} which uses Voronoi tessellations to provide the most `round' bins. The two techniques applied to our NGC 4696 data are shown in Fig. \ref{binning}.

\begin{figure*}
\centering
\includegraphics[width=0.3\textwidth]{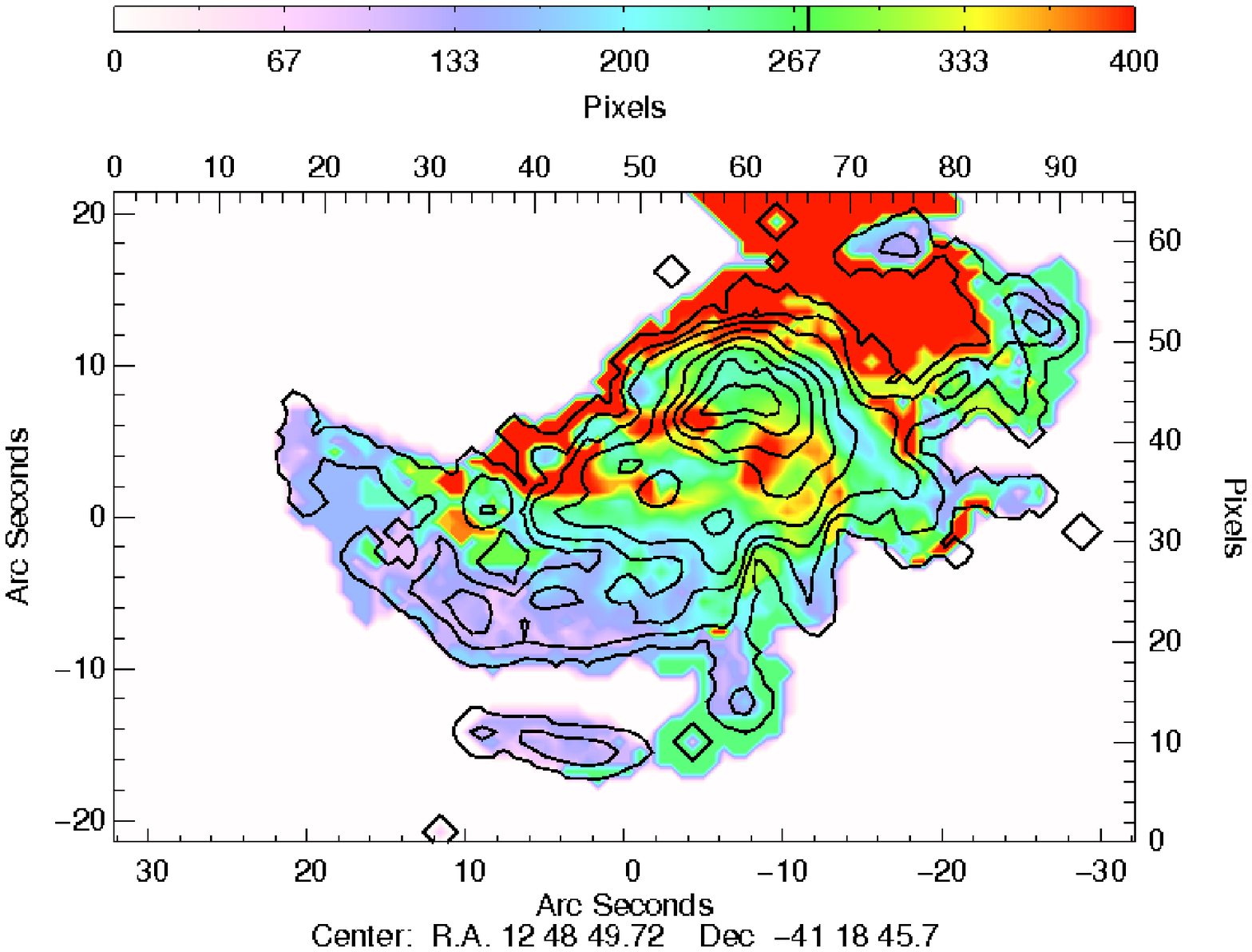}
\includegraphics[width=0.3\textwidth]{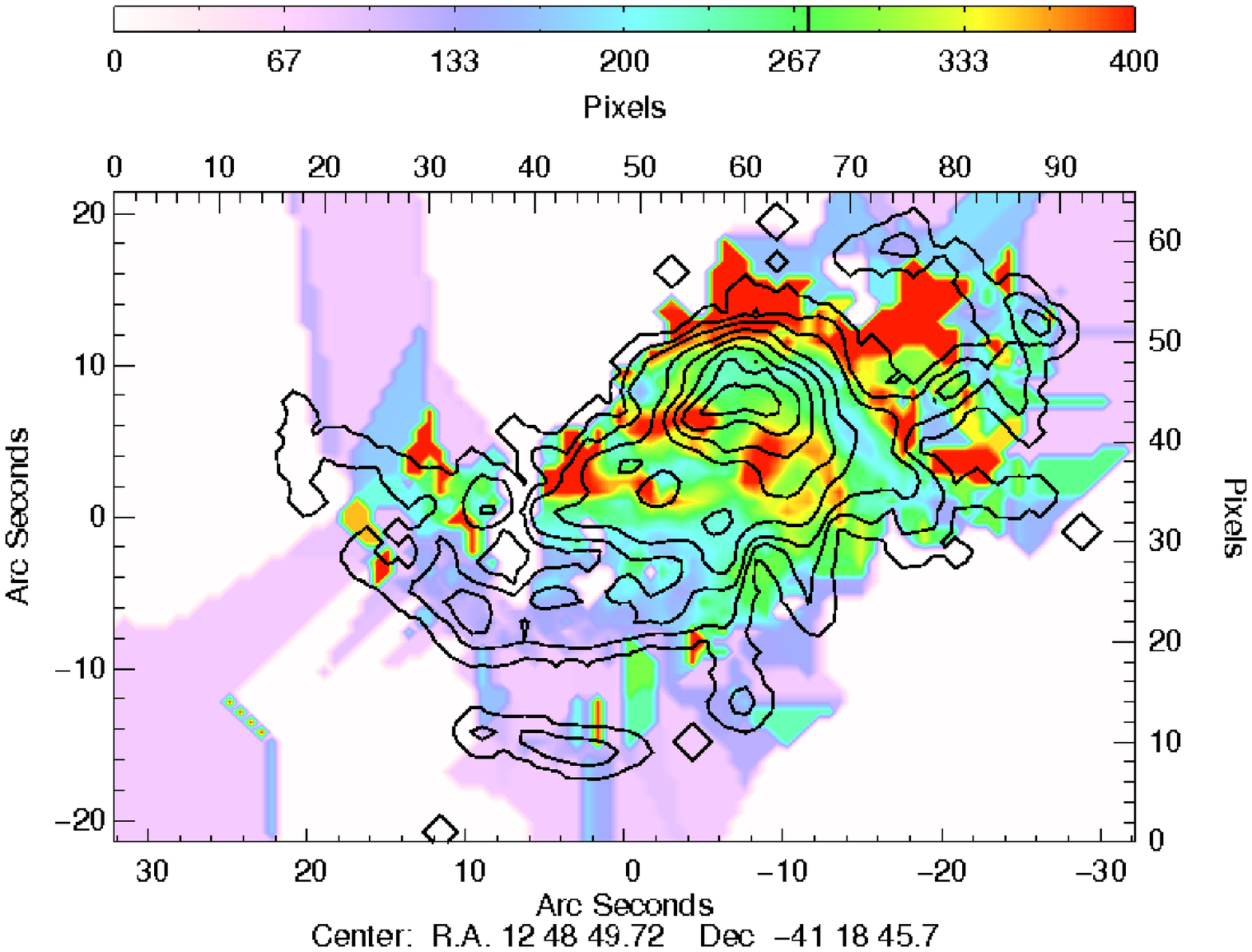}
\includegraphics[width=0.3\textwidth]{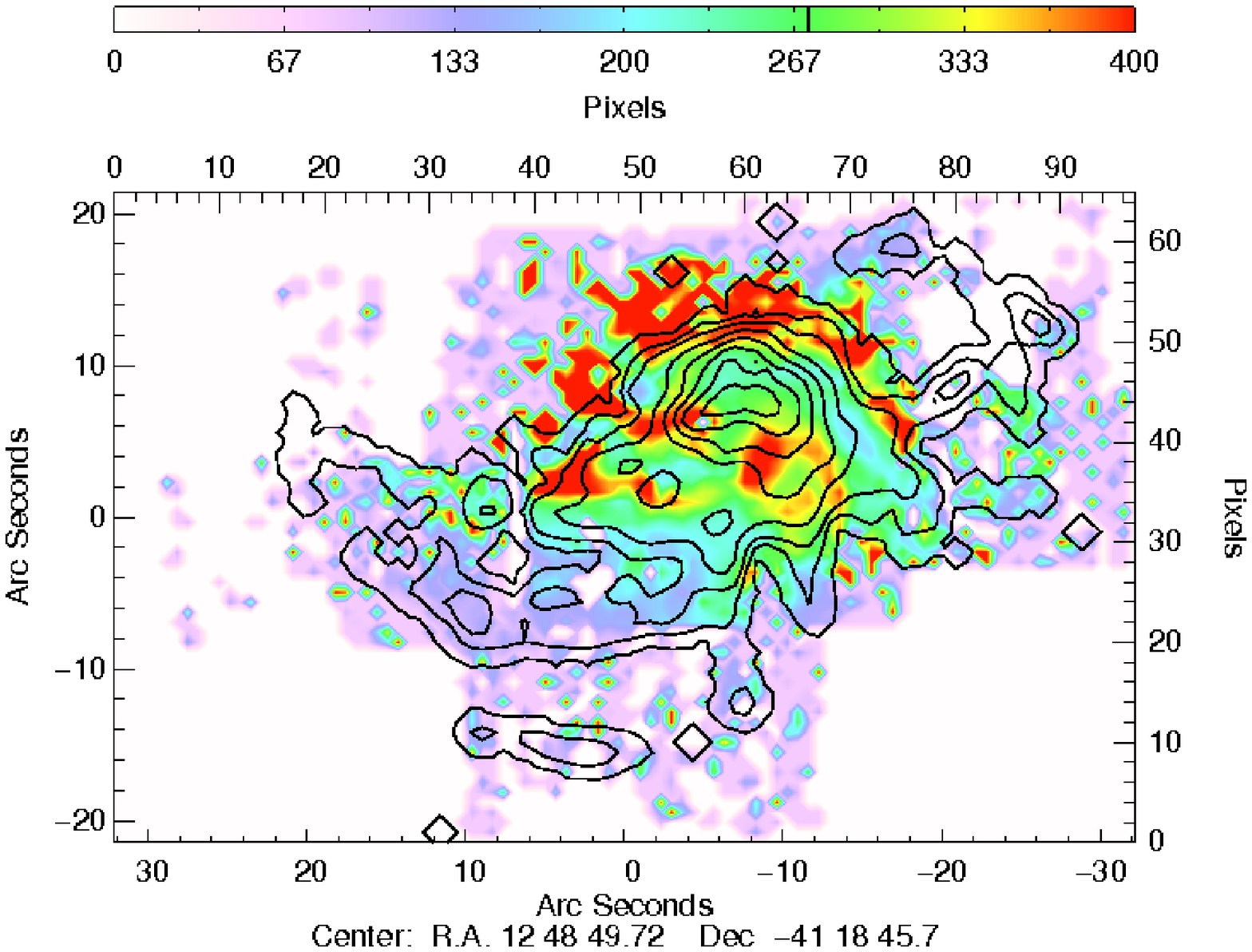}
\caption{FWHM velocity dispersion, in units of \kmps, for different binning techniques for a single 
gaussian fit to the [N {\small II}], H$\alpha$ and [S {\small II}] lines. Left:
Binning to a signal-to-noise of 10 using the contour binning technique of 
\protect \cite{sanders2006c}. Middle: Binning to a signal-to-noise of 10 
using the Veronoi tessellation technique of \protect \cite{cappellari2003}. Right:
The velocity dispersion on a per-pixel basis across the field of view. Here the (0,0) position corresponds to the centre of the image. \label{vel_bins}}
\end{figure*}

Both techniques give similar fits over comparable regions giving us confidence that the physical 
properties of the regions we are binning are similar to the limit of our
spatial resolution (see Fig. \ref{vel_bins}). Due to the filamentary structure
of the optical line nebulosity surrounding NGC 4696 we choose to bin our spectra
using the contour binning technique of \cite{sanders2006c}. 

\subsection{Velocity component fitting}

The data are fit with multiple velocity components. An example single velocity component fit to the H$\alpha$, [N {\sc ii}] and [S {\sc ii}] emission lines are shown in Fig. \ref{fit1}. An F-test is then performed to determine whether extra velocity components are required. 

\begin{figure*}
\centering
\includegraphics[width=0.32\textwidth]{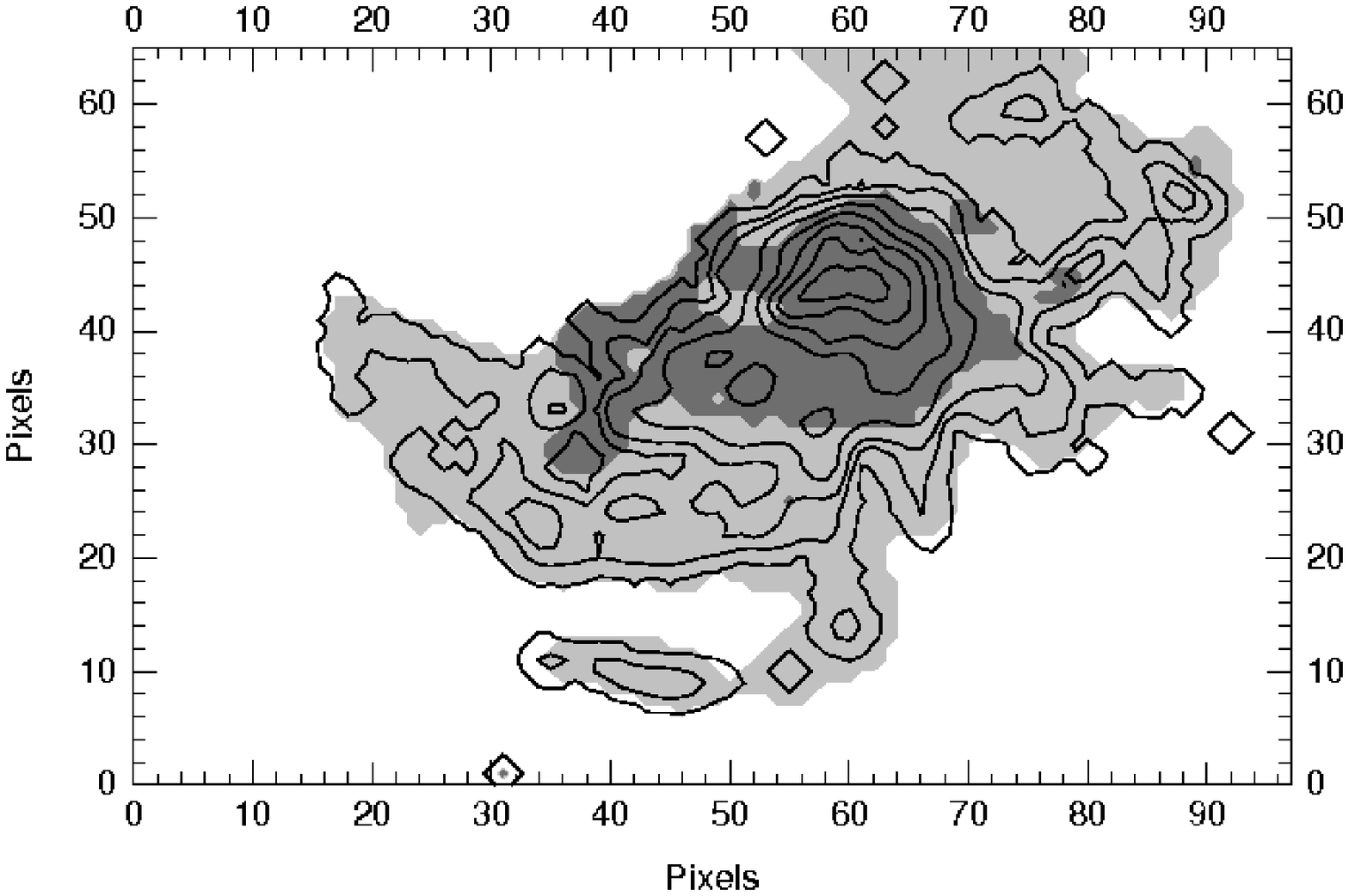}
\includegraphics[width=0.32\textwidth]{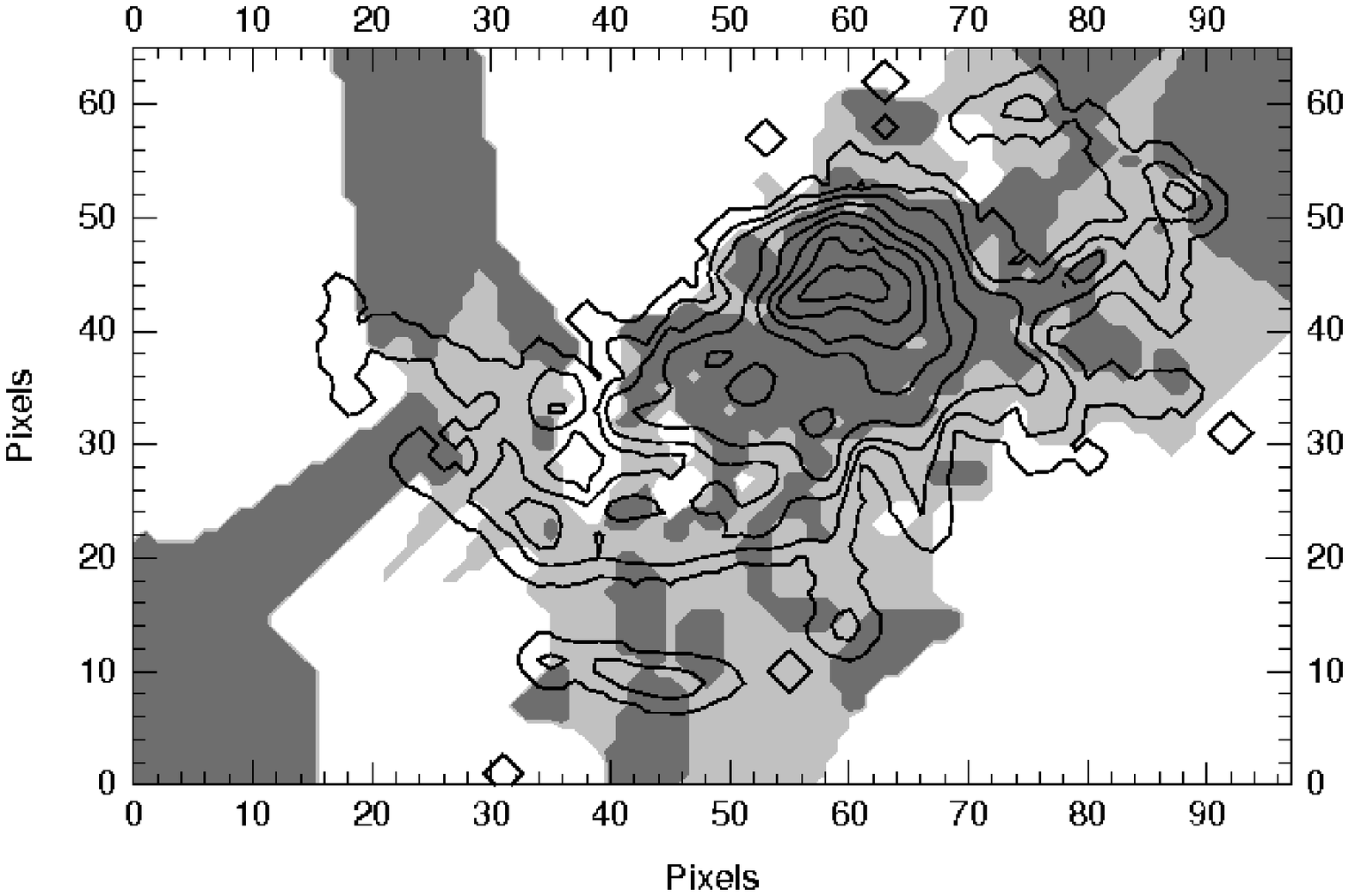}
\includegraphics[width=0.32\textwidth]{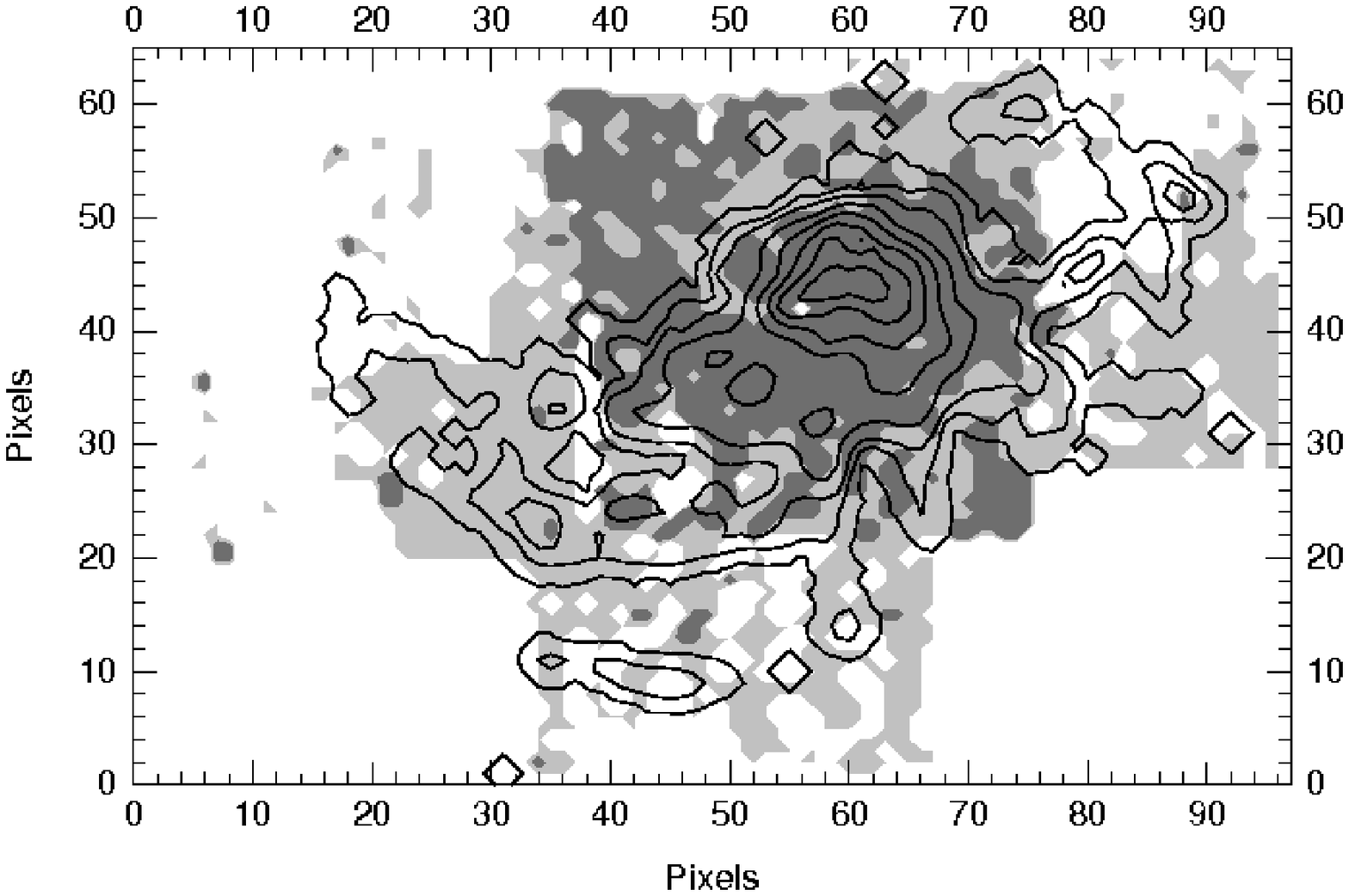}
\caption{Regions where the fibres are statistically found to 
be described by one or two velocity components for the spectra binned by surface brightness (left), 
Voronoi binning (middle) and no-binning (right). The darker shade indicates two components were necessary to fit 
the spectra. Regions in a lighter shade were found to be modelled sufficiently with a single
velocity component. \label{ftest2}}
\end{figure*}

\begin{figure}
\centering
\includegraphics[width=0.45\textwidth]{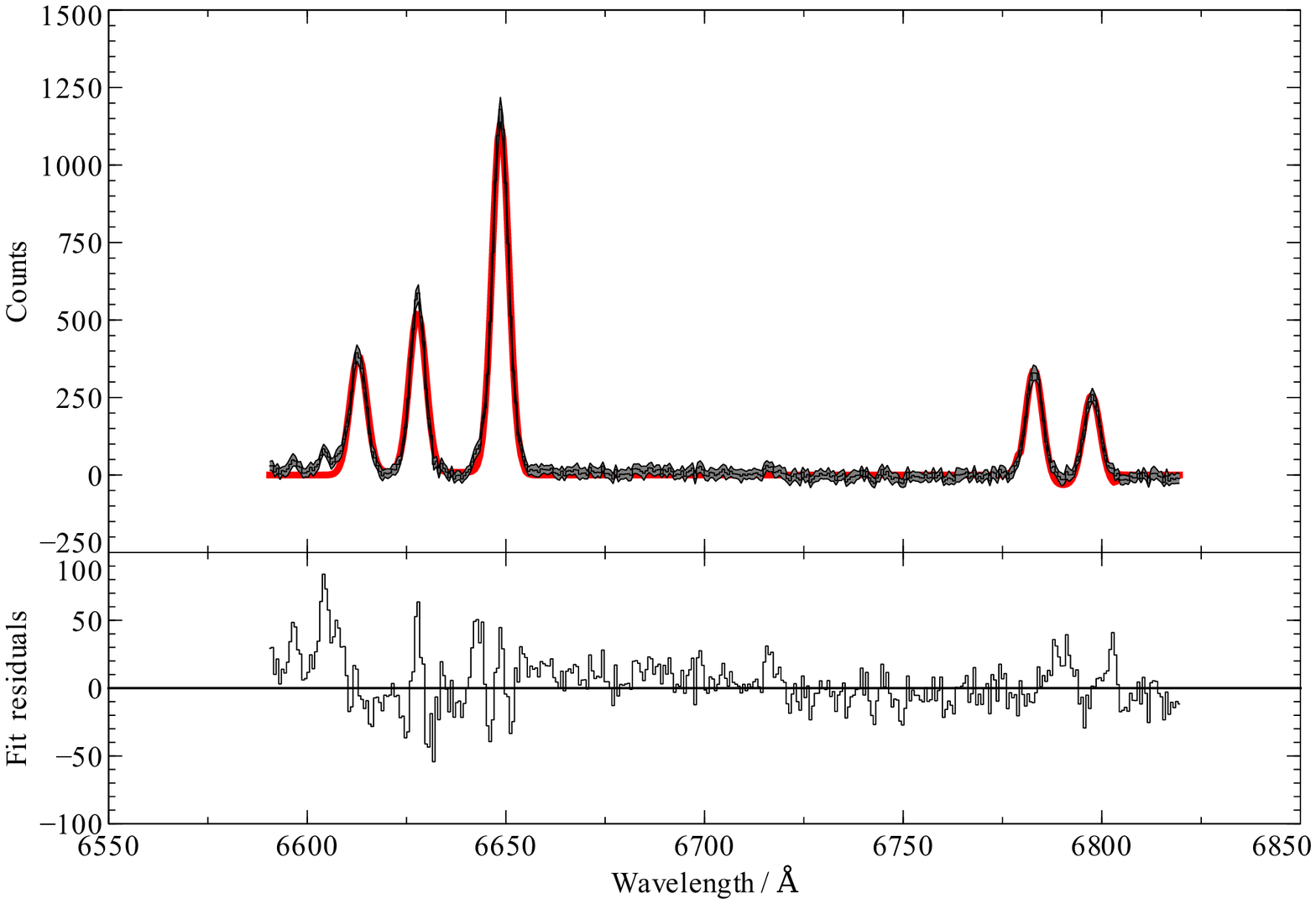}
\caption{An example single gaussian fit to the strongest emission lines in our spectra. These are the 
lines of [N {\small II}]$\lambda$6548, H$\alpha$ $\lambda$6563, [N {\small II}]$\lambda$6583,
[S {\small II}]$\lambda$6717 and [S {\small II}]$\lambda$6730. These lines are
fit between 6590 and 6820~$\mathrm{\AA}$. A fixed continuum has been subtracted and
is estimated as the median value between 6700$-$6750~$\mathrm{\AA}$. The continuum regions
around the H$\alpha$+[N {\small II}] doublet and the [S {\small II}] doublet are fit
separately. The data and error are shown as a thick grey line, the red line indicates the model fit to the data and the fit residuals are shown in the bottom panel. \label{fit1}}
\end{figure}

The lower panels in Fig. \ref{ftest2} show the results of the F-test when the data
is binned by surface brightness contour binning, Voronoi tessellations and no-binning. The darker shade
implies two velocity components are necessary, the light shade indicates the spectra
were sufficiently well represented with a single gaussian velocity component.

\section[]{Emission line maps}
\label{appendixb}

\begin{figure*}
\centering
\includegraphics[width=0.8\textwidth]{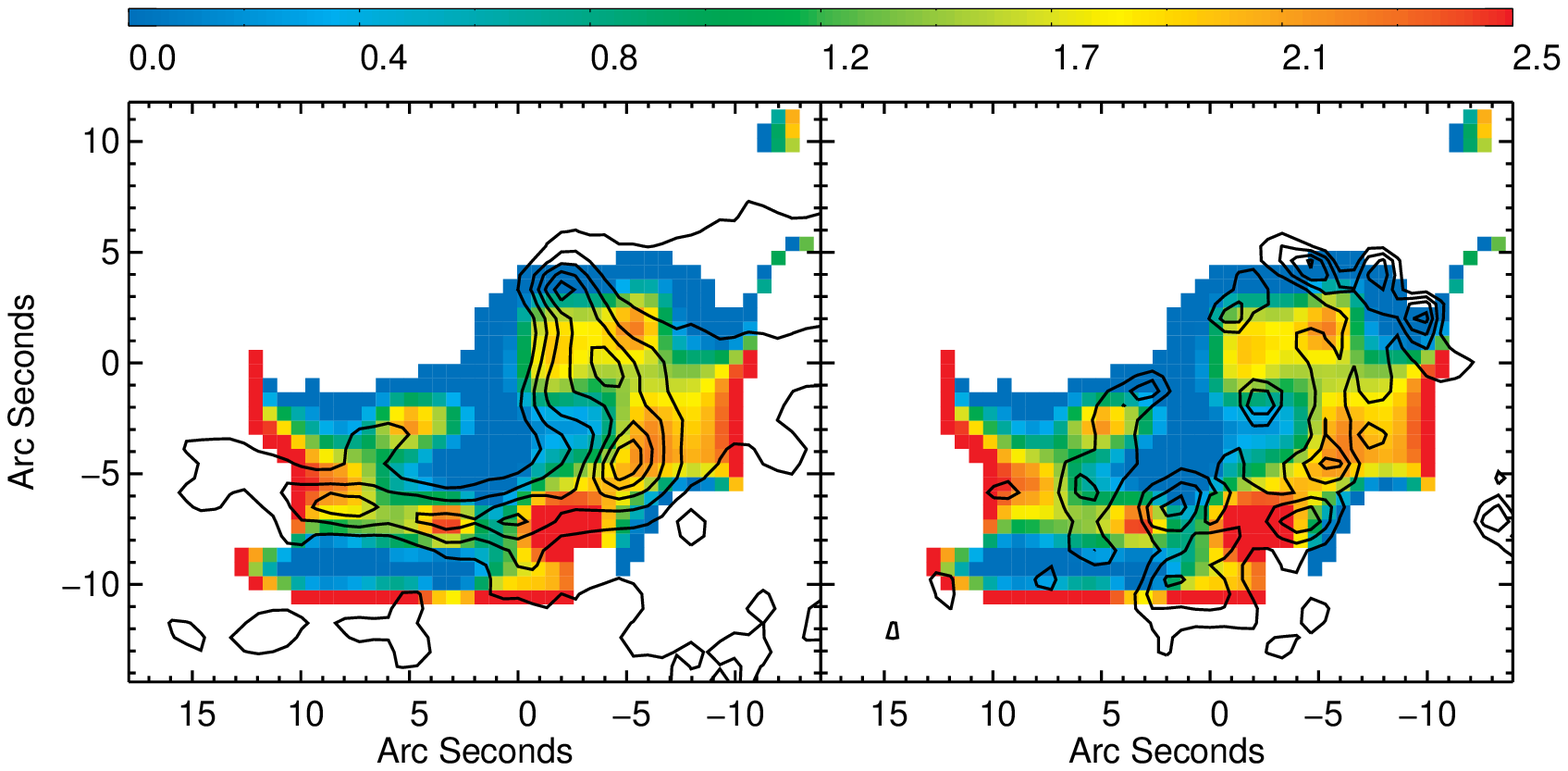}
\caption{
$A_{V}$ intrinsic extinction maps derived from the H$\alpha$/H$\beta$ ratio, fit with a single gaussian. Left: Overlaid with contours of E(B-V) Right: Overlaid with contours of X-ray N$_{\mathrm{H}}$. Contours are from \protect \cite{crawford2005}. The colour bar shows the intrinsic $A_{V}$ extinction calculated assuming case B \citep{osterbrock2006}; only pixels where both H$\alpha$ and H$\beta$ were detected at 4 sigma or above are shown, the original fits were smoothed by a gaussian with FWHM 3 pixels (2'').
\label{reddening}}
\end{figure*}

The reddening map for a one component velocity fit, on a per pixel basis, in the central regions of NGC 4696 is shown in Fig. \ref{reddening}. Contours of E(B-V) derived from HST B and I band images and \chandra\ X-ray N$_{\mathrm{H}}$ column density \citep{crawford2005} are overlaid. 

\begin{figure*}
\centering
\includegraphics[width=0.33\textwidth]{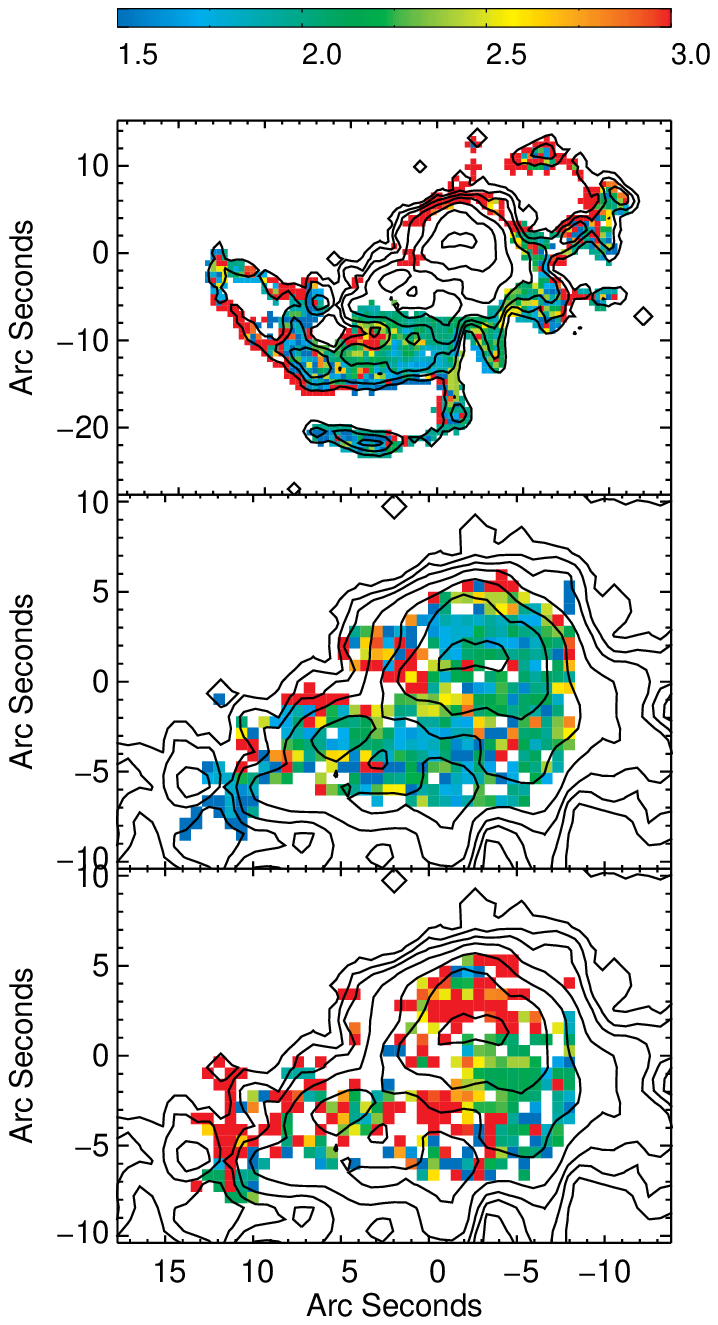}
\includegraphics[width=0.33\textwidth]{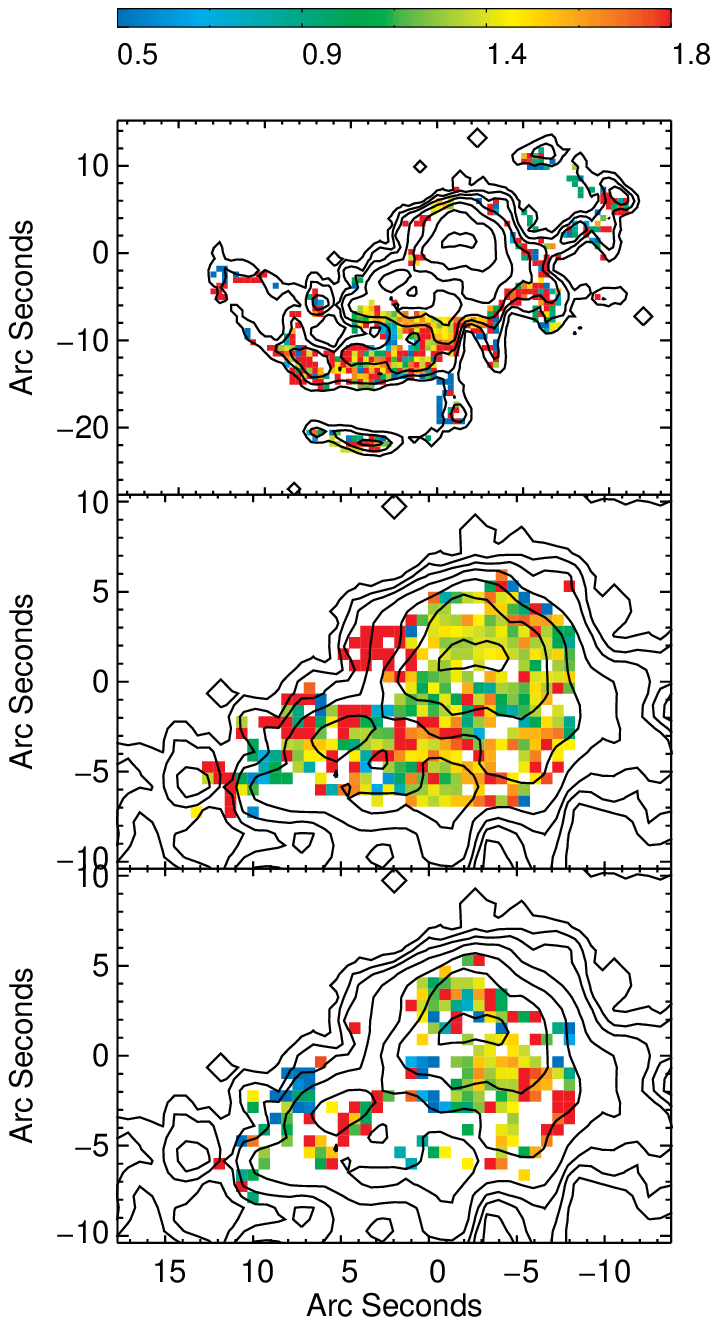}
\includegraphics[width=0.33\textwidth]{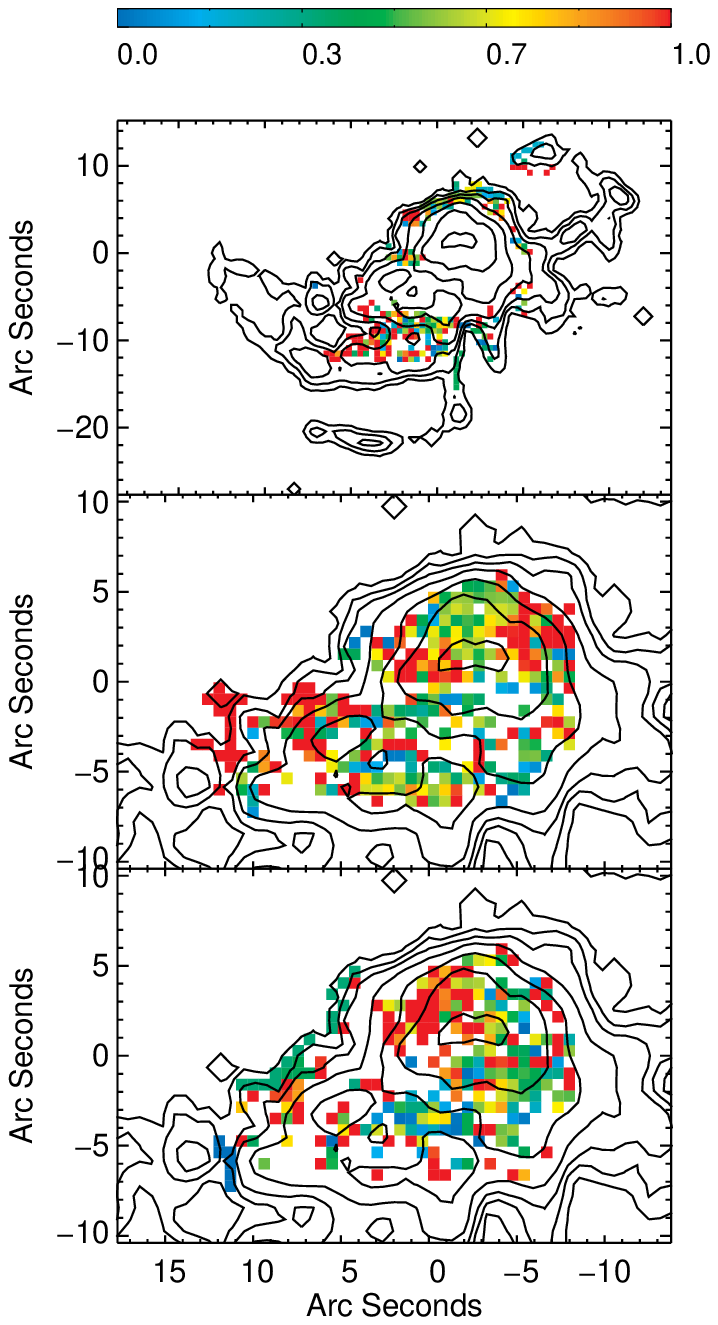}
\includegraphics[width=0.33\textwidth]{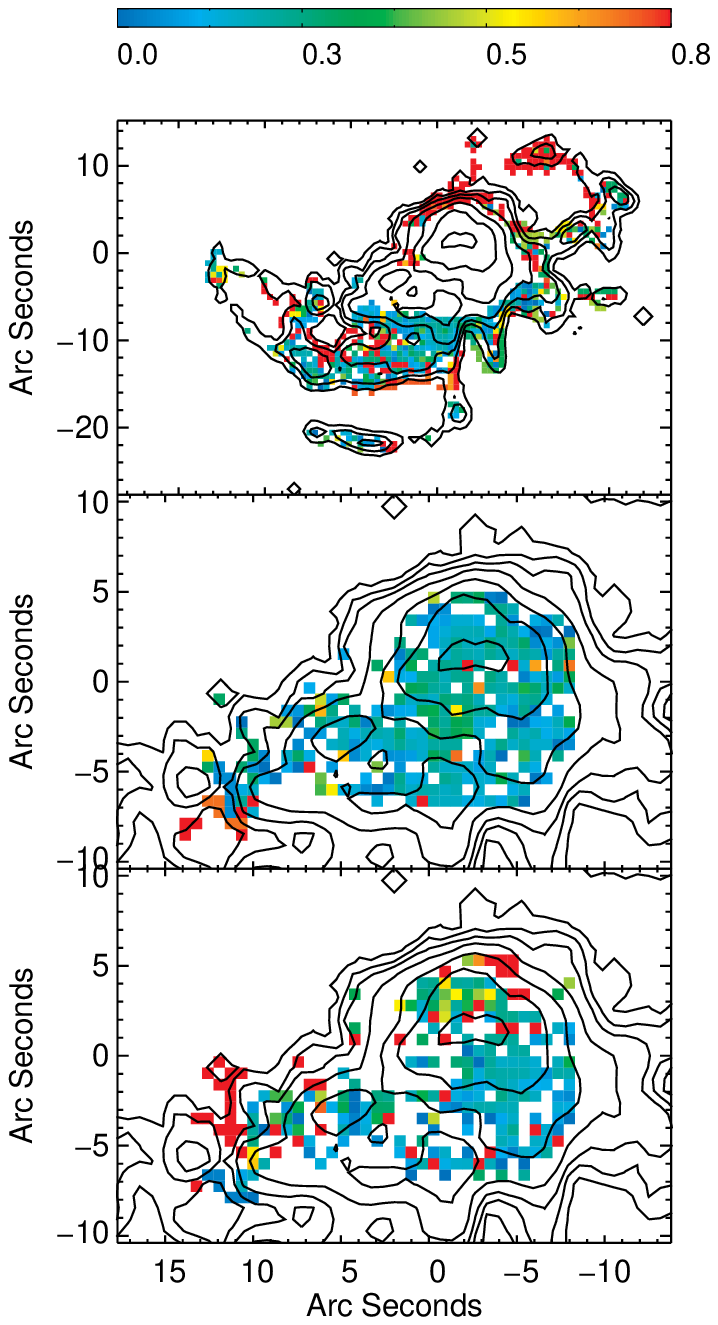}
\includegraphics[width=0.33\textwidth]{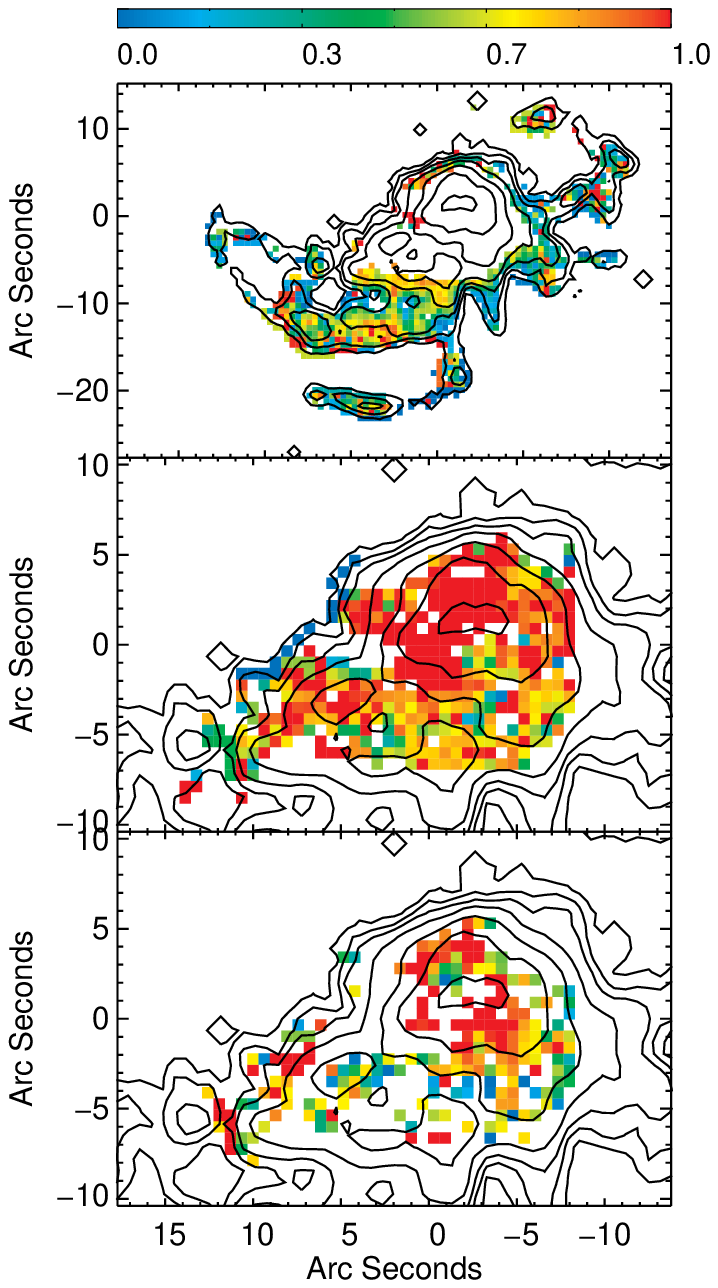}
\caption{Clockwise from top left: [N {\small II}]$\lambda$6583/H$\alpha~\lambda$6563, [S {\small II}]$\lambda$6716/[S {\small II}]$\lambda$6731, [O {\small III}]$\lambda$5007/H$\beta$~$\lambda$4861, [S {\small II}]$(\lambda6716+\lambda6731)$/H$\alpha~\lambda$6563 and [O {\small I}]$\lambda$6300/H$\alpha~\lambda$6563 emission. The top panel in each figure is the ratio in the outer filaments where only one velocity component was detected. The middle panel is the narrow component of the two component gaussian fit and the bottom panel is the broad component. \label{ratios}}
\end{figure*}

Maps of our emission line ratios, binned using the contour binning algorithm of \cite{sanders2006c} are shown in Fig. \ref{ratios}. The top panel showing the single gaussian fits, the middle showing the broad central velocity component and the bottom panel in each case shows the narrow component. Only pixels where the line emission is detected at greater than 4$\sigma$ are shown. The [O {\sc iii}]]/H$\beta$ ratios cannot be traced across the whole length of the outer filaments as our HRB data only cover the central 27 arcseconds.

\end{document}

%% file: defn.tex
\newcommand{\Mpc}{\rm\thinspace Mpc}
\newcommand{\kpc}{\rm\thinspace kpc}

\newcommand{\km}{\rm\thinspace km}

\newcommand{\cm}{\rm\thinspace cm}

\newcommand{\Myr}{\rm\thinspace Myr}
\newcommand{\s}{\rm\thinspace s}

\newcommand{\Msun}{\hbox{$\rm\thinspace M_{\odot}$}}

\newcommand{\keV}{\rm\thinspace keV}

\newcommand{\erg}{\rm\thinspace erg}

\newcommand{\ergpcmsqps}{\hbox{$\erg\cm^{-2}\s^{-1}$}}

\newcommand{\ergpcmsqpsparcsecsq}{\hbox{$\erg\cm^{-2}\s^{-1}\mathrm{arcsec}^{-2}$}}

\newcommand{\ergps}{\hbox{$\erg\s^{-1}\,$}}

\newcommand{\kmps}{\hbox{$\km\s^{-1}\,$}}

\newcommand{\kmpspMpc}{\hbox{$\kmps\Mpc^{-1}\,$}}

\newcommand{\asec}{\rm\thinspace arcsec}

\def\Starlight{\hbox{\sc STARLIGHT}}

\def\chandra{{\it Chandra}}

